\documentclass[12pt]{article}

\usepackage{jheppub}

\usepackage{subcaption}
\usepackage{xcolor}
\usepackage{amsmath}
\usepackage{amssymb}
\usepackage{amscd}
\usepackage{dsfont}
\usepackage{enumerate}
\usepackage{enumitem}
\usepackage{amsfonts}
\usepackage{epsfig}
\usepackage{mathtools}
\usepackage{yfonts}
\usepackage{bbold}
\usepackage{breqn}
\usepackage[utf8]{inputenc}
\usepackage[english]{babel}
\usepackage[subpreambles=true]{standalone}
\usepackage{import}
\usepackage{caption}
\usepackage{multirow}
\usepackage{cases}

\def\s{{\it s}}

\def\eqa{\begin{eqnarray}}
\def\eqae{\end{eqnarray}}
\def\eq{\begin{equation}}
\def\eqe{\end{equation}}
\def\be{\begin{equation}}
\def\ee{\end{equation}}
\def\bea{\begin{eqnarray}}
\def\eea{\end{eqnarray}}
\def\ba{\begin{array}}
\def\ea{\end{array}}

\title{Subsystem entropy  in 2d CFT and KdV ETH}
\author[a,d]{Liangyu Chen,} 
\author[b]{Anatoly Dymarsky,} 
\author[a,c]{Jia Tian,}
\author[a]{and Huajia Wang}
\affiliation[a]{Kavli Institute for Theoretical Sciences, University of Chinese Academy of Sciences, Beijing 100190, China}

\affiliation[b]{Department of Physics and Astronomy, University of Kentucky, Lexington, 40506}
\affiliation[c]{State Key Laboratory of Quantum Optics and Quantum Optics Devices, Institute of Theoretical Physics, Shanxi University, Taiyuan 030006, P.~R.~China}
\affiliation[d]{Yau Mathematical Sciences Center, Tsinghua University, Beijing 100084, China}

\emailAdd{liangyu-chen@mail.tsinghua.edu.cn}
\emailAdd{a.dymarsky@uky.edu}
\emailAdd{wukongjiaozi@ucas.ac.cn}
\emailAdd{wanghuajia@ucas.ac.cn}

\abstract 
{
We study subsystem  entropy in 2d CFTs, for subsystems  constituting  a finite fraction of the full system. We focus on the extensive contribution, which scales linearly with the subsystem size in the thermodynamic limit. We employ the so-called diagonal approximation to evaluate subsystem entropy for the chaotic CFTs in thermal state (canonical ensemble), microcanonical ensemble, and in a primary state, matching previously known results. We then proceed to find analytic expressions for the subsystem entropy at leading  order in $c$, when the global CFT state is the  KdV generalized Gibbs ensemble  or the KdV microcanonical ensemble. 
Previous studies of primary eigenstates have shown that,  akin to  fixed-area states in AdS/CFT, corresponding subsystem entanglement spectrum is flat. This behavior is seemingly in sharp contradiction with the one for the thermal (microcanonical) state, and thus in apparent contradiction with the subsystem Eigenstate Thermalization Hypothesis (ETH). In this work, we resolve this issue by comparing the primary state with the KdV (micro)canonical ensemble. We show that the results are consistent with the KdV-generalized version of the subsystem ETH, in which local properties of quantum eigenstates are governed by their values of conserved KdV charges. Our work
solidifies evidence for the KdV-generalized ETH in 2d CFTs and emphasizes Renyi entropy as a sensitive probe of the reduced-density matrix. 

}

\begin{document}
\maketitle
\section{Introduction}\label{sec:intro}
Subsystem entropy has become an indispensable tool to study and characterize quantum systems. This is particularly true for two-dimensional conformal field theories, due to seminal works of Calabrese and Cardy \cite{Calabrese:2004eu,Calabrese:2009qy} and subsequent developments \cite{Calabrese:2005in,Casini:2006es,Calabrese:2007mtj,Calabrese:2009ez,Calabrese:2010he,Casini:2008wt,Casini:2009sr}. These works influenced intensive study of entanglement entropy in other contexts, in particular for one-dimensional lattice systems \cite{alba2009entanglement,Vidmar:2017pak,Eisler_2017,Lydzba:2020qfx,PhysRevB.99.075123,leblond2019entanglement,PhysRevE.105.014109,PhysRevE.107.064119,rodriguez2023quantifying,PhysRevB.108.245101,Cheng:2023omo}. In holography, study of entanglement led to celebrated Ryu-Takayanagi proposal \cite{Ryu:2006,Ryu:2006extend} and the subsequent works \cite{Fursaev:2006ih, HRT:2007, Nishioka:2009un, Headrick:2010, Casini:2011, Hung:2011ta, Hung:2011, Wall:2012, Hartman:2013, Faulkner:2013renyi, Hartnoll:2013, Chen:2014unl, Lewkowycz:2013nqa, Faulkner:2013qc, Chen:2013dxa, Perlmutter:2014, Engelhardt:2014g,Asplund:2015,Dong:2016renyi,Dong:2016n1, Bianchi:2016n2,Dong:2016HE,Rangamani:2016, fixedarea2}. More recently this progress resulted in a breakthrough in understanding the entropy of Hawking radiation \cite{Penington:2019npb,Almheiri:2019psf,Almheiri:2019hni}.

Our goal in this paper is to use the subsystem entropies in 2d  CFTs to probe the entanglement structure of different ensembles, including individual CFT eigenstates, and shed light on the Eigenstate Thermalization Hypothesis (ETH) \cite{Deutsch:1991,Srednicki:1994,Rigol:2008,Alessio:2016} in these theories. More recent and refined understanding of ETH  claims that matrix elements in quantum ``chaotic''  systems admit a statistical description \cite{Pappalardi:2022aaz}, giving rise to matrix model universality \cite{PhysRevLett.123.260601,Jafferis:2022uhu,Wang:2023qon}. Subsystem entropy alone is not powerful enough to probe these aspects of ETH, but can be used to verify the ``diagonal part'' of the subsystem ETH \cite{Dymarsky:2018}, which states that individual energy eigenstates, reduced to a sufficiently small subsystem $A$ exhibit universality -- reduced state is determined by thermodynamically relevant (conserved) quantities, i.e.~in general case it is a smooth function of energy
\bea
\rho_A^i={\rm Tr}_{\bar A}|E_i\rangle \langle E_i|,\qquad \rho_A^i\approx \rho_A(E_i).
\eea
Using state-operator correspondence, one can show that in 2d CFTs a version of subsystem ETH follows from the  conventional ``local'' ETH, which requires expectation value of all local primary operators $O$ in highly excited  eigenstates $|E_a\rangle$  to match  thermal expectation values, $\langle E|O|E\rangle = O_{\rm th}(T(E))$,  up to exponentially-small corrections  \cite{Lashkari:2016vgj}.

Entanglement entropy was used to probe ETH in 2d CFTs  early on, for the subsystems smaller than the inverse temperature. In this limit the relation between subsystem entropy and expectation values of local operators is explicit \cite{BinChen:2013,Lashkari:2014yva,Lashkari:2015dia,Lin:2016,Sarosi:2016oks}. A direct comparison of  Renyi entropy for a primary eigenstate and a thermal state revealed a discrepancy at leading order in large central charge, suggesting conventional ETH in 2d CFTs is broken \cite{He:2017,Basu:2017kzo, He:2017txy,Guo:2018}.   This apparent breakdown of ETH is  further solidified by a mismatch of Renyi entropy for large subsystems, also at  leading order in large $c$ \cite{Wang:2018}. By a large subsystem, here and in what follows, we understand the subsystem which scales as a fixed fraction of the full system in the thermodynamic limit.

It is well-known that  in full generality 2d CFTs admit an intrinsic integrable structure: an
infinite tower of mutually-commuting charges ${{Q}}_{2k-1}$ \cite{Bazhanov:1994KdV,Bazhanov:1996KdV,Bazhanov:1998KdV},  which includes the  CFT Hamiltonian $H={{Q}}_{1}$. These so-called KdV charges are thermodynamically relevant;  they are conserved during  time evolution, and therefore an equilibrium state emerging dynamically from an initial out-of-equilibrium configuration would be labeled by the  original values of ${{Q}}_{2k-1}$. Following the ideas of \cite{Rigol2007} it was proposed in \cite{Cardy:2015xaa} that in 2d CFTs the emerging equilibrium state would be in the form of KdV Generalized Gibbs Ensemble (KdV GGE). Following on this idea, it was further proposed in \cite{Lashkari:2017hwq,He:2017txy} that 2d CFTs instead of conventional ETH satisfy generalized ETH  in the sense of \cite{PhysRevLett.106.140405,Ilievski_2015,Vidmar_2016}, with the reduced density matrix being a smooth function of  ${\mathcal{Q}}_{2k-1}$, matching reduced KdV GGE state in the appropriate limit, 
\bea
\begin{aligned}
     \rho_A^i={\rm Tr}_{\bar A}|E_i\rangle \langle E_i|,\qquad \rho_A^i\approx \rho_A({{Q}}^i_{2k-1}), \qquad \hat{{Q}}_{2k-1} |E_i\rangle={{Q}}^i_{2k-1}|E_i\rangle.
\end{aligned} 
\eea
This proposal, which we will refer to as KdV ETH, was verified and established for local operators from the identity block, and hence for reduced states $\rho_A$ when the subsystem $A$ is small and CFT is chaotic, up to first {\it non-trivial} order in $1/c$ in \cite{Dymarsky:2019etq}.\footnote{For the quasi-primaries from the identity family conventional ETH is satisfied only at leading order in $1/c$ \cite{Basu:2017kzo}.
This leads to a mismatch between primary state and thermal state Renyi entropies already at leading order in $c$ \cite{Lashkari:2016vgj,Basu:2017kzo, He:2017txy,Guo:2018}.} In this paper we further support the KdV ETH proposal by showing that it successfully reproduces large subsystem Renyi entropy results for the canonical and microcanonical ensembles \cite{Calabrese:2009qy, Dong:2018} and for  the primary eigenstates \cite{Wang:2018}, and reconciles the mismatch between them. 

The subsystem ETH alone is not sufficient to predict the functional form of von Neumann or Renyi entanglement entropy of a subsystem, except that it would be a  smooth function of total energy (and other conserved charges in the generalized ETH case). When the subsystem $A$ is large, it is a natural expectation that the structure of the boundary terms at $\partial A$ do not  affect the entropy density of $A$ at leading order.  This assumption gives rise to semiclassical diagonal approximation of $\rho_A$, when the diagonal terms dominate and are given by \cite{Dymarsky:2018},
\bea
\label{semic}
\langle E_A|\rho_A(E)|E_A\rangle ={\Omega_{\bar A}(E-E_A)\over \Omega(E)}.
\eea
Here  $|E_A\rangle$ is an eigenstate of the subsystem $A$ and $\Omega, \Omega_{\bar A}$ are densities of states of the full system and the complement of $A$, which we call $\bar A$. We will elaborate on this formula later in the paper. For CFTs in any dimensions,  the semiclassical approximation \eqref{semic} readily reproduces the volume part of the subsystem von Neumann and Renyi entropies, when the full system is in the Gibbs (canonical) or microcanonical ensembles. When the the full system is in an eigenstate, we will see that \eqref{semic} is still sufficient for describing higher-dimensional CFTs, but in 2d it has to be modified by taking KdV charges into account. 

Our main goal will be to employ the  diagonal approximation that includes the KdV charges, and apply it to compute Renyi entropy when the full system is in the KdV GGE or KdV microcanonical ensemble. We compare these results with those for a primary eigenstate, which exhibits an index-independent  subsystem Renyi entropy $S^n_A$ when $A$ is large. This implies the corresponding  entanglement spectrum that is flat, and  the distribution of eigenvalues of the reduced density matrix is  $\delta$-function peaked. We find that while this behavior is in sharp contrast with that one of the  ordinary (micro)canonical  ensemble, it is consistent with and emerges as a limit of the KdV (micro)canonical ensemble, with the values of charges that match those of the primary state. The agreement between the primary state  and the KdV-charged ensembles provides a new consistency check for the 2d CFT KdV ETH proposal. This is one of the two  main messages of this paper. Second message is the  power of the diagonal approximation, which allows us  to evaluate subsystem entropy of various global ensembles. 

The Eigenstate Thermalization Hypothesis is only expected to emerge in the thermodynamic limit,  e.g.~when the full system size is substantially larger than the typical length scale associated with the energy/charge densities of the considered states. For CFTs one can rescale the total system size to be one, when  thermodynamic limit becomes the high energy limit. Alternatively there could be another regime of validity, when the central charge, which controls the number of local degrees of freedom is large, $c\gg 1$. In this paper, most of the discussion, including validity of the semiclassical diagonal approximation, relies on the high energy limit. Certain results presented in this paper are valid only in  large $c$ limit, which  is taken in addition to the high energy (i.e.~thermodynamic) limit. Such a scenario of large $c$ and large energy  arises naturally for holographic CFTs. We mention here  the companion paper \cite{KdVETHholography}, where we discuss the computation of the subsystem entropy for holographic theories from the bulk perspective.

The paper is organized as follows. We start in next Section \ref{sec:setup} by reviewing the general CFT setup. The discussion there  also includes a brief overview of quantum KdV hierarchy in 2d case. In section \ref{sec:ETH} we remind the reader  how the ETH in CFT is formulated in the general case and the regimes when it is expected to apply.  We then proceed in Section \ref{sec:semiclassical} with the  review of   semiclassical diagonal approximation  and how to incorporate KdV charges in the 2d case. Section \ref{sec:main} discusses large subsystem entropy  when the full state is in one of the following states: canonical and microcanonical ensembles, a heavy primary state, and also KdV generalized Gibbs ensemble and 
KdV-generalized  microcanonical ensemble.  We conclude with a discussion in Section \ref{sec:discussion} . 

\section{CFT preliminaries and setup}
\label{sec:setup}
Throughout the paper we consider 2d CFTs on a cylinder: the spatial circle of circumference $L$ times physical (Minkowski) time. Such theories would emerge e.g.~as an effective description in the continuous limit of 1D periodic gapless spin-chains. Field theory at finite inverse temperature $\beta$ would be described by the Euclidean 2d CFT on a torus with the modular parameter $\tau=i\beta/L$. Using conformal symmetry  we can rescale $L$ to be $2\pi$,  $L$ can be always  restored  using dimensional analysis. It is related to the parameter $\ell$ of   \cite{Dymarsky:2018lhf,Dymarsky:2018iwx} as follows, $L=2\pi \ell$.

The CFT  Hilbert space is a direct product of left and right sectors, such the Hamiltonian is a direct sum
\bea
{\ell\, H}={\mathcal{Q}}_{1}+\bar{\mathcal{Q}}_{1}=\left(L_0-{c\over 24}\right)+\left({\bar L}_0-{c\over 24}\right),
\eea
and the eigenstates $|E_L,E_R\rangle=|E_L\rangle \otimes |E_R\rangle$. To keep the notations concise we will only explicitly write the left sector, e.g.~the Hamiltonian
\bea
H\equiv Q_1={\mathcal{Q}_1\over \ell},\quad {\mathcal{Q}}_{1}=\int_0^{2\pi} {d\varphi\over 2\pi}\, T=L_0-{c\over 24},
\eea
with the eigenvalues $E=(\Delta+n-c/24)/\ell$, where $\Delta,n$ are the primary dimension and the descendant level correspondingly. 

The CFT density of states (this is the density of states of left sector only) admits the following universal form given by Cardy formula \cite{CardyFormula}
\bea
\label{Omega2d}
\Omega_L(E)=e^{S(E,L)},\qquad S(E,L)=\sqrt{c\pi LE/3}.
\eea
This expression applies either when $E L\gg 1$ or $c\gg 1$ \cite{Hartman:2014oaa}. As we mentioned in the Introduction,  these are two regimes when we expect ETH to apply. We mainly focus on the  regime $E\rightarrow \infty$, which corresponds to  thermodynamic limit when the system size $L\rightarrow \infty$, while energy density $e=E/L$ is kept fixed, i.e.~fixed effective temperature $T^2=\beta^{-2} \propto E/L$. For 2d CFTs, once $L$ is re-scaled to be $2\pi$, thermodynamic limit corresponds to vanishing inverse temperature and large energy,
\bea
\beta\propto L^{-1},\quad  \Delta+n \propto  c\, \beta^{-2}, 
\eea 
where $L\rightarrow \infty$ is now a formal dimensionless parameter. 

Written in the form \eqref{Omega2d}, Cardy formula simply states that in the  large volume limit the entropy $S$ is extensive and the effective temperature dependence is fixed by dimensional analysis 
\bea
S(E)=\kappa\, L\, T, \qquad {dS\over d E}=\beta\equiv T^{-1},  \label{Cardy}
\eea 
such that the only non-trivial content of Cardy formula is the relation between the coefficient  $\kappa={c\pi/6}$ in \eqref{Cardy} and CFT central charge $c$. 

The Cardy formula allows various improvements. General CFT state $|\Delta+n\rangle$ is a descendant of a primary $\Delta$ of level $n$. The density of the descendants can be understood combinatorially. The density of primaries, beyond $\Delta \rightarrow \infty$ or $c\rightarrow \infty$ regimes, is not universal but satisfies various consistency conditions \cite{Mukhametzhanov:2018zja,Mukhametzhanov:2020swe}. The corrections to Cardy formula are not important for our consideration, as we are only interested in the extensive part of the subsystem entropy in the thermodynamic limit, the part that scales linearly with the volume, and it is not sensitive to these non-universal corrections. 

Our main focus is on ``large'' subsystems, consisting of a single interval of length $L_A$, that is a finite fraction of the total length $L$. We assume that $0< x\equiv L_A/L\leq 1$ is kept fixed in the thermodynamic limit $L\rightarrow \infty$. In this limit we expect subsystem Renyi entropy $S^n_A$ to be proportional to the subsystem size $L_A$
with the  entropy density being a function of { intrinsic} (local) properties $\zeta$ of the CFT  state $\rho$, fractional parameter of the subsystem size $x$, and the  Renyi index $n$. We thus write
\bea
\label{calS}
S_A^n= L\,{\cal S}(n,x,\zeta)+o(L_A), 
\eea
where ${\cal S}/x$ is the entropy density. Function $\cal S$ is linear in $x$ for small $x\ll 1$. We prefer to work with ${\cal S}$ instead of entropy density ${\cal S}/x$,  because  $x$-dependence of $\cal S$ defines the Page curve. In what follows,  while we explicitly do calculations only in the chiral sector, we write $\cal S$ for the full theory assuming ${\bar c}=c$.

It is helpful to compare this setting with the ``small'' subsystem limit, mentioned in the Introduction. It corresponds to the limit of fixed $L_A$ and fixed temperature $T=\beta^{-1}$ in thermodynamic limit $L\rightarrow \infty$. This regime usually assumes the  quantization scheme where the CFT is defined on a plane, but also can be obtained as a limit of carefully taking $x\rightarrow 0$ while $L$ tends to infinity. When the combination $L_A T$ is small, subsystem entropy can be evaluated in terms of expectation values of local operators \cite{BinChen:2013,Lashkari:2014yva,Lashkari:2015dia,Lin:2016,Sarosi:2016oks,He:2017,Basu:2017kzo, He:2017txy,Guo:2018}, reducing the  study of entropy in an eigenstate to ``local ETH'' \cite{Lashkari:2016vgj}.

The case of higher dimensional CFTs is similar. There is no left-right decomposition of Hilbert space, and $E$ stands for full energy. We consider the $d+1$ theory quantized on a $d$-dimensional spatial manifold $\cal M$ of volume $V$, which is taken to infinity. In this limit the density of states $\Omega(E)=e^S$ is controlled by entropy $S$, which is extensive
\bea
\label{Shd}
S(E)=\kappa V T^d \propto (V E^d)^{1/(d+1)}.
\eea
``Large'' subsystem limit assumes that the ratio $x=V_A/V$ and inverse temperature $\beta$ are fixed while $V\rightarrow \infty$. After  rescaling of $V$, thermodynamic limit corresponds to the high energy limit $\beta\propto L^{-1}$,  $E\propto  L^{d+1}$, where $L\rightarrow \infty$ is a formal dimensionless parameter.   

The crucial difference between the 2d and higher dimensional cases is that the ``stress-energy sector'' of 2d CFTs is universal and exhibits an intrinsic integrable structure. As a result 2d theories have an infinite tower of conserved KdV charges \cite{Bazhanov:1994KdV,Bazhanov:1996KdV,Bazhanov:1998KdV}, constructed from the stress tensor  and its derivatives. The first few charges are 
\bea
\begin{aligned}
     {\mathcal{Q}}_{1} = \int^{2\pi}_0 \frac{d\varphi}{2\pi} T,\quad {\mathcal{Q}}_{3} = \int^{2\pi}_0 \frac{d\varphi}{2\pi} (TT), \quad
     {\mathcal{Q}}_{5} = \int^{2\pi}_0 \frac{d\varphi}{2\pi} \left(T(TT)+\frac{c+2}{12}(\partial T)^2\right),\nonumber
\end{aligned}
\eea
also see \cite{Dymarsky:2019iny} for  a detailed discussion of ${\mathcal{Q}}_{7}$. These  operators are made of the generators of the Virasoro algebra, they commute with each other, and hence with the Hamiltonian. They map descendants of a given primary $\Delta$ and given level $n$ into each other. From the dimensional analysis, when $\Delta, n$ are large, we have that ${\mathcal{Q}}_{2k-1}$ scales as ${{\mathcal{Q}}}_{1}^k$. After $L$ is  restored
\bea
Q_{2k-1}=(2\pi)^{k}{\mathcal{Q}_{2k-1}\over L^{2k-1}},
\eea
charges $Q_{2k-1}$ become extensive \cite{Dymarsky:2018lhf}.  For a  2d CFT on a unit circle conventional thermodynamic limit  assumes the following scaling 
\bea
\label{KdVthermol}
\beta\propto L^{-1},\quad  {\mathcal{Q}}_{2k-1}  \propto  c^k  L^{2k}, 
\eea 
where $L\rightarrow \infty$ is a formal dimensionless parameter. Finally, the KdV charges ${\mathcal{Q}}_{2k-1}$ and their densities $q_{2k-1}\equiv Q_{2k-1}/L$  satisfy the following positivity conditions
\bea
q_{2k-1}\geq q_1^k. \label{qineq}
\eea

\section{ETH in CFT: a brief overview}\label{sec:ETH}
To put things into perspective, in this section we provide a brief overview of the Eigenstate Thermalization Hypothesis (ETH) in the context of CFTs. Understood narrowly, ETH postulates that the individual energy eigenstates are thermal, in the sense of expectation values of  a physically-motivated set of suitable operators \cite{Deutsch:1991,Srednicki:1994,Rigol:2008,Alessio:2016},
\bea
\langle E|O|E\rangle=O^{\rm th}(T),\qquad T=T(E).  \label{ETHdef}
\eea
For this to be true, it is necessary and sufficient for $\langle E|O|E\rangle$ to be a smooth function of the energy density $E/V$, smoothness is defined as finite derivative in the thermodynamic limit $V\rightarrow \infty$. This is the so-called ``diagonal part'' of ETH. This condition can be formulated without specifying an observable, in terms of the reduced state of a subsystem $\rho_A={\rm Tr}_{\bar A}|E\rangle \langle E|$; this formulation is called the subsystem ETH \cite{Dymarsky:2018}. A more complete statement of the ETH would refer to statistical properties of matrix elements, and also include off-diagonal matrix elements \cite{Srednicki:1994}. We focus on the diagonal part in this paper.

In the case of $d>2$ CFTs, the algebraic relation between primary and descendant states preclude ETH in its straightforward form \eqref{ETHdef}. It was proposed in \cite{Lashkari:2016vgj} that in chaotic CFTs \eqref{ETHdef} would be satisfied  only  by primary operators $O$ and primary states $|E\rangle$, a condition which can be reformulated in terms of the heavy-heavy-light OPE coefficients \cite{Lashkari:2016vgj, Lashkari:2017hwq}
\bea
C_{HHL}\propto {\rm H}^{{\rm L}/(d+1)}, \label{ETHCFT}
\eea 
where the proportionality coefficient is fixed by a dimensionless combination involving  thermal expectation value of light operator $\langle O_L\rangle_T  T^{-L}$ and $\rm H,L$ in \eqref{ETHCFT} are dimensions of heavy and light operators correspondingly.  The ETH prediction \eqref{ETHCFT} for CFTs remains a conjecture which, to date, was not observed or verified.\footnote{The behavior \eqref{ETHCFT} is guaranteed {\it on average}, the non-trivial part is the prediction that heavy-heavy-light OPE coefficient will be sufficiently smooth for ${\rm H}>>1$, with asymptotically vanishing deviations from the mean value. One can hope that with time this behavior could be probed e.g.~for the 3D Ising model using numerical conformal bootstrap \cite{PhysRevD.86.025022} or using Hamiltonian truncation \cite{Fitzpatrick:2022dwq}.} For 2d CFTs, thermal expectation values of primaries vanish in the thermodynamic limit $\beta \to 0$, because thermal cylinder is conformally flat. The ETH   then predicts that $C_{HHL}\rightarrow 0$ when  ${\rm H}\rightarrow \infty$. 
Besides the diagonal part, one also expects the ``off-diagonal'' OPE coefficients involving heavy primary operators with distinct conformal dimensions to exhibit statistical properties dictated by the full statements of the ETH. This idea, which we would refer to as OPE ETH, is actively explored for 2d CFTs in the context of holography \cite{Kraus:2016nwo,Brehm:2018ipf,Romero-Bermudez:2018dim,Chandra:2022bqq}.

In 2d CFTs, vanishing of heavy-heavy-light OPE coefficients would ensure vanishing of all diagonal matrix  element \eqref{ETHdef} with $O$ being a primary or a descendant light operator and primary or descendant high energy states $|E\rangle$. This behavior is expected for 2d chaotic  theories.\footnote{What defines a chaotic CFT is somewhat ambiguous, but the holographic CFTs \cite{Hartman:2014oaa} are expected to fall into this category.}
It is necessary for the diagonal part of ETH to hold, but not quite sufficient. This is because the identity operator and its descendants always have non-vanishing eigenstate and thermal expectation values. To ensure the diagonal part of ETH, expectation values of quasi-primaries should be thermal, but this is simply not the case already  at first subleading order in $1/c$, as was shown in e.g.~\cite{Basu:2017kzo}. It was proposed in \cite{Lashkari:2017hwq} and later established in \cite{Dymarsky:2019etq} that a {\it generalized} version of ETH is taking place in 2d, when the expectation values in an eigenstate are smooth functions of the eigenstate KdV charges
\bea
\label{KdVETH}
\langle E|{ O}|E\rangle={ O}(Q_{2k-1}).
\eea
In what follows we refer to \eqref{KdVETH} and  its subsystem version as the KdV ETH. 

The KdV ETH only applies to operators (quasi-primaries) $ O$ from the identity block, expectation values of all other operators are expected to vanish  in high energy eigenstates because of vanishing OPE coefficients. 
The expectation values \eqref{KdVETH} for the vacuum block quasi-primaries are fixed by the Virasoro algebra. They are essentially theory-independent, the only dependence is through central charge $c$ and  quantum numbers of the state $|E\rangle$ (including corresponding primary dimension $\Delta$). A smooth, in fact polynomial form of $ { O}(Q_{2k-1})$ was established in \cite{Dymarsky:2019etq} at first two orders in $1/c$  expansion; further $1/c$ corrections or finite $c$ behavior in thermodynamic limit when $E\rightarrow \infty$ is an open question \cite{inprogress}.

In this paper we study the subsystem version of the KdV ETH. This involves checking the relation between subsystem reduced density matrices of high energy eigenstates and the KdV generalized thermal ensembles carrying the same KdV charges. Regarding the subsystem density matrix $\rho_A$, the von Neumann entropy 
\be
S^{vN}_A = -\text{Tr}\, \rho_A \ln{\rho_A}
\ee
is a useful probe that quantifies the amount of entanglements between the subsystem and its complement. However, more refined properties of the reduced density matrix, such as the full entanglement spectrum, can be recovered from  the subsystem Renyi entropy for all integer $n>1$,
\be
\label{Rentropy}
S^n_A = \frac{1}{1-n} \ln{\text{Tr}\,\rho^n_A}. 
\ee

The subsystem version of the  ETH \eqref{ETHdef} would ensure that the subsystem Renyi entropy of an eigenstate (when the subsystem is less than the half system size) is the same as for the microcanonical ensemble \cite{Dymarsky:2018}. Large subsystem Renyi entropy of a primary high energy eigenstate was evaluated in \cite{Wang:2018}, where it was shown to deviate from the microcanonical result already at the leading order in $c$. This was another stark evidence that the conventional ETH does not apply in 2d CFT case. To probe subsystem KdV ETH in this paper we evaluate Renyi entropy for the KdV generalized microcanonical ensemble and show it reproduces the primary state result in the appropriate limit. Thus, our results provide a  consistency check of the subsystem KdV ETH proposal.

\section{Subsystem entropy  and semiclassical diagonal approximation}\label{sec:semiclassical}
In this section we explain the semiclassical diagonal approximation of \cite{Dymarsky:2018} and generalize it for the case when a system admits a tower of conserved charges. 
The main idea behind this approximation  is that when a large system is split into a large subsystem $A$ and its (also large) complement, the boundary conditions at $\partial A$ should not affect subsystem entropy at leading  (extensive) order.\footnote{This seemingly natural expectation, which is obvious for classical
systems in Gibbs ensemble, is much more non-trivial  in case of  subsystem Renyi entropy for quantum systems even those exhibiting finite correlation length \cite{Renyi}.} Taking this idea to extreme, we can decouple subsystem $A$ from its compliment $\bar A$ by considering the Hamiltonian: 
\be \label{eq:Hamiltonian_additive}
H=H_A+H_{\bar A}
\ee 
In principle there should be additional terms $\Delta H$ encoding interactions between subsystems $A$ and $\bar{A}$. For local Hamiltonians we expect such terms to be sub-dominant in the thermodynamic limit with both subsystem sizes $L_A$ and $L_{\bar{A}}$ scaling linearly with $L$. We therefore ignore such terms and treat (\ref{eq:Hamiltonian_additive}) at face value. From here we readily obtain an expression for the reduced density matrix of an eigenstate $|E\rangle$ of $H$, written in the eigenbasis $|E_A\rangle$ of $H_A$ \eqref{semic}. It can be approximated by the diagonal form
\bea
\begin{aligned}
     \rho_A(E) ={\rm Tr}_{\bar A}|E\rangle \langle E| 
     = \int dE_A \left[{\Omega_{\bar A}(E-E_A)\over \Omega(E)}\right] |E_A\rangle \langle E_A|. \label{diagonalap}
\end{aligned}
\eea
Going back to original Hamiltonian, which includes interactions between $A$ and $\bar A$, in the limit when both $A$ and $\bar{A}$ are large, at leading order one expects the full density of states $\Omega$  to be the convolution of $\Omega_A$ and $\Omega_{\bar A}$. In this limit we expect 
$\rho_A$ to be diagonally dominated with \eqref{diagonalap} providing an approximation sufficient to evaluate subsystem entropy at leading (volume-law) order \cite{Dymarsky:2018}. We remark that the validity of the semiclassical diagonal approximation (\ref{diagonalap}) is controlled essentially by the suppression of the interaction term $\Delta H$ relative to that of $H_A$ and $H_{\bar{A}}$. For states with energy densities characterized by the effective length scale $\beta$ this corresponds to\footnote{We do not have to assume the original system is 1-dimensional. In general $d$, linear size $L$ would be substituted by the volume $V$ and \eqref{eq:semiclassical_validity} would become $V_A,V_{\bar A}\gg \beta^d$.}  
\be\label{eq:semiclassical_validity}
L_A, L_{\bar{A}} \gg \beta. 
\ee
This condition is automatically satisfied in the thermodynamic limit. 

The discussion above implicitly assumes a spatially-extended lattice system, with local interactions described by $H$ and the Hilbert space decomposing into ${\cal H}_A \otimes {\cal H}_{\bar A}$. In case of CFTs, the Hilbert space decomposition does not apply, and the subsystem eigenstates $|E_A\rangle$ are not well defined. It would be interesting to develop this picture rigorously by introducing conformal boundary conditions at $\partial A$, such that $H_A$ and $H_{\bar A}$ would describe boundary CFTs. In what follows, we will take a different approach by thinking of CFTs as emerging in a continuous limit from lattice systems, where $|E_A\rangle$ are well-defined. 

In order to obtain closed-form expression for the subsystem entropy, one  needs an explicit expression for the subsystem density of states $\Omega_A(E_A)$ and $\Omega_{\bar{A}}\left(E_{\bar{A}}\right)$. For large subsystems, in the thermodynamic limit $E\gg 1/L$, we expect that the subsystem entropy density at leading order is the same as  thermal entropy density $\s$, which is a functions of energy density. We therefore propose that at leading order 
\be\label{eq:entropy_ansatz}
\ln \Omega_A(E_A) \approx  {S\left(E_A,L_A\right)},\quad \ln\Omega_{\bar{A}}(E_{\bar{A}}) \approx { S\left(E_{\bar{A}},L_{\bar A}\right)},
\ee
where  thermal entropy  $S(E,L)$ of the full system at global energy $E$ is given by \eqref{Omega2d}. It is convenient to introduce energy density $e=E/L$ and similarly $e_A,e_{\bar A}$ for subsystems  such that 
\bea
\label{se}
S(E,L)=L\,s(e), \;\ {\s}(e)=\sqrt{c\pi e/3},\;\ e=E/L,
\eea
and
\bea
\begin{aligned}
     \ln \Omega_A=L\, x\, {\s}(e_A)+o(L),\quad
     \ln \Omega_{\bar A}= L\, (1-x)\, {\s}(e_{\bar A})+o(L),
\end{aligned}
\eea
where  $x = L_{A}/L$  and $o(L)$ indicate sub-extensive corrections. 

To evaluate subsystem entropy in a global ensemble characterized by certain energy density distribution, one would first integrate \eqref{diagonalap}  with an appropriate measure over $e$ to obtain corresponding subsystem density matrix, and then integrate its power over $e_A$ to evaluate \eqref{Rentropy}. In the thermodynamic limit $L\rightarrow \infty$ both integrals are given by saddle point approximation.  A number of explicit examples of Renyi entropy calculation using diagonal approximation will be given in the next section. 

This approach evaluating subsystem entropy was further developed in 
\cite{Huang:2017std,Grover:2017,PhysRevE.100.022131}, by postulating  eigenstate decomposition 
\bea \label{decomposition}
|E\rangle= \sum C^E_{ij}|E_A^i \rangle \otimes |E_{\bar A}^j\rangle
\eea 
with random coefficients $C_{ij}^E$. This approximation  does not do justice to 
 local operators located at the boundary of $A$, leading to their incorrect expectation values in state $|E\rangle$, but yields the same diagonal approximation to subsystem entropy as \eqref{diagonalap}. In systems with local interactions actual  $C^E_{ij}$ exhibit intricate structure, recently explored in \cite{Shi:2023czc}. 

When the system admits additional extensive conserved quantities, \eqref{diagonalap} should be modified as follows. The density of states $\Omega$ is now a function of the vector of conserved charges $\vec{Q}$ and therefore 
\bea
\begin{aligned}
     \rho_A(\vec{Q})={\rm Tr}_{\bar A}|\vec{Q}\rangle \langle \vec{Q}|  
     = \int dQ_A \left[{\Omega_{\bar A}(\vec{Q}-\vec{Q}_A)\over \Omega(\vec{Q})}\right] |\vec{Q}_A\rangle \langle \vec{Q}_A|,
\end{aligned}
\label{diagonalQ}
\eea
where $\vec{Q}$ are the charges of the eigenstate $|\vec{Q}\rangle$, and 
$\vec{Q}_A$ are the charges of the subsystem eigenstate $|\vec{Q}_A\rangle$ (we assume the latter can be defined). 
Although not necessary for what follows, similarly to \eqref{decomposition} one can assume there is an eigenstate decomposition 
\be \label{eq:decomposition_KdV}
|\vec{Q}\rangle = \sum_{\vec{Q}_A,\vec{Q}_{\bar{A}}}C(\vec{Q},\vec{Q}_A,\vec{Q}_{\bar{A}}) |\vec{Q}_A\rangle \otimes |\vec{Q}_{\bar{A}}\rangle,
\ee 
with the function $\;C(\vec{Q},\vec{Q}_A,\vec{Q}_{\bar{A}})$ sharply peaked around $\vec{Q}=\vec{Q}_A+\vec{Q}_{\bar A}$ due to  charge conservation. 

In the thermodynamic limit, at leading order the density  of states should be an extensive function of charge densities $q_i={Q_i/L}$,
\bea
\ln \Omega(\vec{Q})=L\, {s}(\vec{q})+o(L),
\eea
and similarly for subsystems 
\bea
\begin{aligned}
     \ln \Omega_A (\vec{Q}_A )=L\, x\, {s}(\vec{q}_A)+o(L), \quad
     \ln \Omega_{\bar{A}} (\vec{Q}_{\bar{A}} )=L\, (1-x)\, {s}(\vec{q}_{\bar A})+o(L),
\end{aligned}
\eea
where $\vec{q}_A=\vec{Q}_A/L_A$ and $\vec{q}_{\bar A}=\vec{Q}_{\bar A}/L_{\bar A}$. As in case of systems without additional conserved quantities discussed above, the subsystem entropy can be evaluated via saddle point approximation. 

The consideration above is general and can be applied to  any translationally-invariant systems with (quasi)local conserved and mutually-commuting charges. 
In many cases, including the case of quantum KdV hierarchy, there are infinitely many charges $Q_i$. In most practical settings  only a finite number of them are kept fixed or specified. Then one integrates over the remaining charges that are free to vary. For example, upon fixing first $m$ KdV charges ${ Q}_{2i-1}$, $i\leq m$, corresponding density of states would be given by 
\be
\label{intQmany}
\Omega\left(Q_1,...,Q_{2m-1}\right) = \int \prod^\infty_{k=m+1} dQ_{2k-1}\; \Omega(\vec{Q}).
\ee  
In the thermodynamic limit this the integral can be evaluated via saddle point approximation, yielding
\bea
\label{sps}
s(q_1,\dots, q_{2m-1})=\left. s(\vec{q})\right|_{q_{2i-1}^*},
\eea
where $q_{2i-1}^*$ for $i> m$ is defined such  that $\left.\partial s/\partial q_{2i-1} \right|_{q_{2i-1}^*}=0$.
In particular, by fixing only density of first KdV charge (energy density) $e\equiv q_1$ and integrating over $q_{2i-1}$ for $i\geq 2$, we should recover conventional CFT density of states given by Cardy formula with $s(e)$ given by \eqref{se}.

\subsection{KdV microcanonical density of states}
\label{SQ1Q3}
In general the KdV microcanonical density of states $s(\vec{q})$, which would count the total number of CFT states $\Omega= e^{L\,s+o(L)}$ with the given values of KdV charges $Q_{2i-1}=L\,q_{2i-1}$ is not known. In what follows we focus on the case when the density of only lowest two charges $q_1,q_3$ are specified and discuss  the behavior of $s(q_1,q_3)$. 

First, instead of $q_3$ it is convenient to introduce non-negative variable $\epsilon$, 
\bea
\label{defepsilon}
q_3=q_1^2(1+\epsilon),\quad \epsilon\geq 0. 
\eea
Then the entropy density $s(q_1,q_3)$ can be rewritten as a function of $q_1,\epsilon$. Since the entropy is extensive we can parameterize it as 
\bea
\label{sq1q3}
s(q_1,q_3)=\sqrt{c\pi q_1\over 3}\s(\epsilon,c),
\eea
where $\s(\epsilon,c)$ is a function  of $\epsilon$ and the central charge $c$ only.

Integrating over $q_3$ for any fixed $q_1>0$ should yield back Cardy formula $s=\sqrt{c\pi q_1/3}$, from where we can conclude that $\s(\epsilon,c)$ achieves its global maximum $\s=1$ at 
\be
\label{epsilonstar}
\epsilon^* = \frac{22}{5c}.
\ee
This expression is exact for any value of $c>1$ and follows from the expectation value of $q_3$ in the (micro)canonical ensemble \cite{Dymarsky:2018lhf}.

To obtain qualitative behavior of $s(\epsilon,c)$ we can model KdV-generalized microcanonical ensemble with fixed $Q_1$ and $Q_3$ by a ``refined'' microcanonical ensemble of all CFT descendant states 
\be
\label{Virstate}
|\Delta, \vec{m}\rangle ={L}_{-m_1} ...{L}_{-m_k} |\Delta\rangle,\;\; |\vec{m}| = \sum^k_{i=1}  m_i=n,\;\; m_i \in \mathds{N},
\ee 
with fixed primary dimension $\Delta$ and the descendant level $n$.  The total density of such states is given by Cardy formula, with the number of descendants approximated by  the Hardy-Ramanujan formula 
\be\label{eq:Cardy_corrected} 
\begin{aligned}
    \Omega(\Delta,n) &= \exp{\left(\pi \sqrt{2(c-1)\Delta/3}+\pi \sqrt{2n/3}\right)},\\
\end{aligned}
\ee 
where $\Delta,n \rightarrow \infty$. The relation between $\Delta$ and $n$, which are assumed to be fixed within a narrow interval, and the KdV charges $Q_1,Q_3$ can be obtained by evaluating the  expectation values of the latter in the refined microcanonical ensemble. This calculation was performed in \cite{Dymarsky:2018lhf},\footnote{The original calculation proceeded by finding the exact expression for ${\rm Tr}_\Delta (q^{L_0} Q_3)$, where the trace is over all descendants of $|\Delta\rangle$, $q=e^{-\beta/\ell}$,  and $\beta$ is related to $n$ by a  Legendre transform.} and in the thermodynamic limit yields
\bea\label{eq:averages}
\begin{aligned}
    \langle q_1\rangle_{\Delta,n}=2\pi {\Delta +n\over L^2}, \quad
    \langle q_3\rangle_{\Delta,n}=\langle q_1\rangle^2_{\Delta,n}+(2\pi)^2 {4\Delta n+(2c/5+4)n^2\over L^4},
\end{aligned}
\eea
where both $\Delta$ and $n$ are assumed to scale as $L^2 \rightarrow \infty$. Plugging this back into \eqref{eq:Cardy_corrected} we find  
\bea
\label{eq:KdV_entropy}
\begin{aligned}
    s^{\rm appr}(\epsilon,c) =  c^{-1}\,\Bigg(\sqrt{5\sqrt{1+c\epsilon/10}-5}  
     +\sqrt{(c-1)\left(5+c-5\sqrt{1+c\epsilon/10}\right)}\Bigg).
\end{aligned}
\eea
We stress, this expression is an approximation. It is only qualitatively correct, as it calculates entropy (number of microstates)  of a different ensemble. 
It can be checked to achieve global maximum  $s(\epsilon^*,c)=1$, consistent with the general proposal.
When $\epsilon=0$, which corresponds to primary states with $n=0$, it yields the correct value $s=\sqrt{1-c^{-1}}$. 
The function \eqref{eq:KdV_entropy} is concave. At large $c$ and for $\epsilon\sim c^0$ it can be simplified as follows 
\be
\label{slargec}
s^{\rm appr}(\epsilon,c) = 1- \sqrt{\frac{5\epsilon}{8c}}+...
\ee
From here we conclude that large $c$ correction to $s=1$ is of order $1/\sqrt{c}$, which  is beyond the scope of leading semi-classic (holographic) limit. We discuss the latter in the companion paper \cite{KdVETHholography}. Using bulk computation we show that the same result $s=1+o(c^0)$   holds for any  $\epsilon$ of order one. Simplified version of this argument can be found in the Appendix \ref{app:shol}.

\begin{figure}
    \centering
    \includegraphics[width=0.85\textwidth]{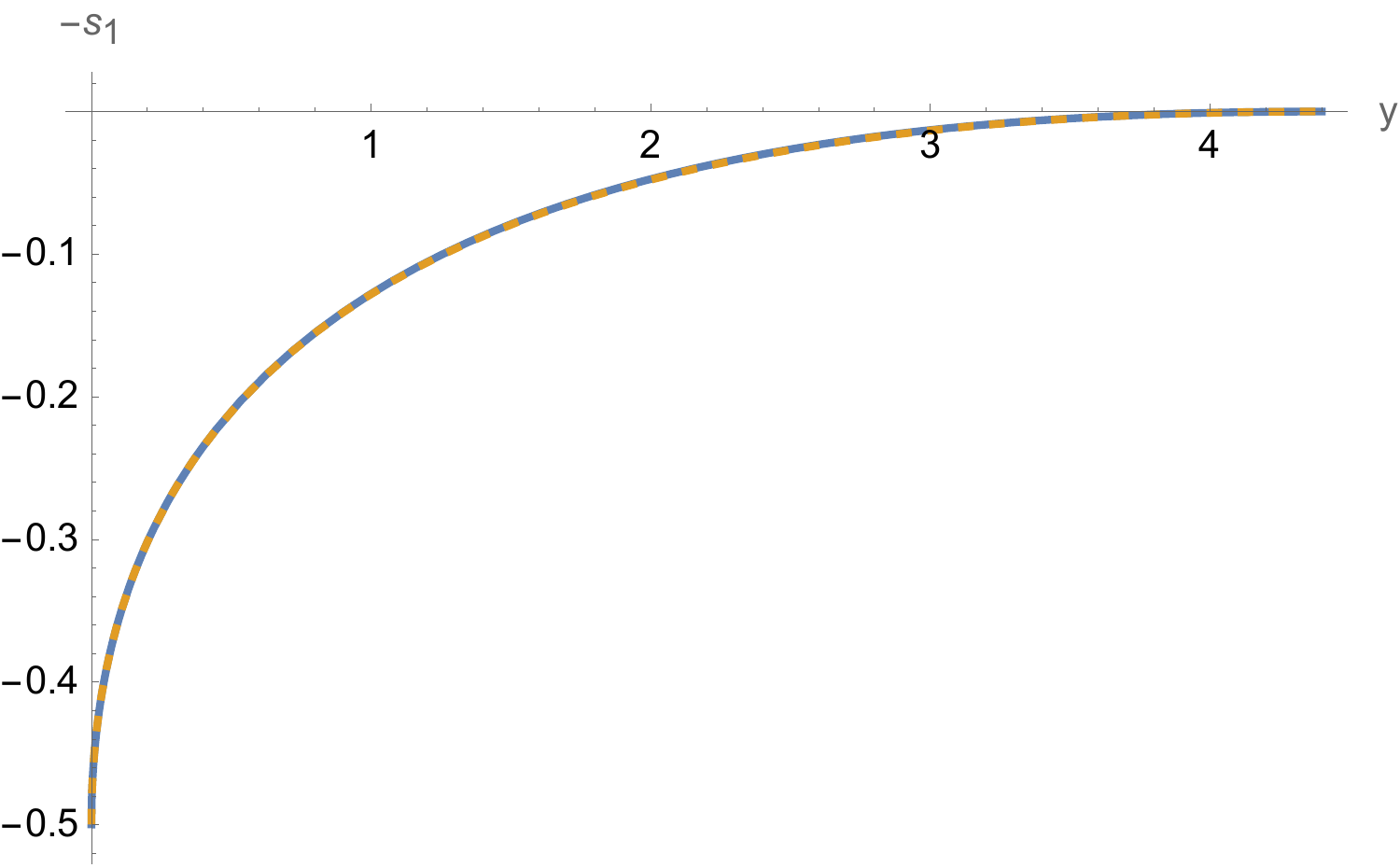}
    \caption{Numerical plot of  $-s_1(y)$ inside  the interval $0\leq y\leq 22/5$ (blue) vs.~the plot of the uncontrolled approximation $-s_1^{\rm appr}(y)$ given by \eqref{naive} (dashed orange).}
    \label{fig:s1}
\end{figure}

Focusing on the interval $0\leq \epsilon \leq 22/(5c)$, it is convenient to introduce $y=c\, \epsilon$ such that  in the large $c$ limit 
\bea
\label{scexp}
s(\epsilon,c)=1-{s_1(y)\over c}+{s_2(y)\over c^2}+\dots
\eea
As follows from \eqref{eq:KdV_entropy} the approximate leading order correction is 
\be
s_1^{\rm appr}(y)=-\frac{1}{2} \sqrt{\frac{5}{2}} \sqrt{y+10}+\sqrt{\sqrt{\frac{5}{2}} \sqrt{y+10}-5}+2. 
\label{naive}
\ee
To calculate exact $s_1$, one can match  the KdV GGE free energy (we absorbed $-\beta$ into the definition of $F$), 
\be
\label{kdvgge13}
\begin{aligned}
    e^{F} &={\rm Tr}\,e^{-\mu_1 Q_1-\mu_3 Q_3} 
     = \int dQ_1\, dQ_3\, \Omega(Q_1,Q_3)\, e^{-\mu_1 Q_1-\mu_3 Q_3},
\end{aligned}
\ee
known up to two orders in $1/c$ expansion, see section \ref{sec:KdVGGE} below.
Starting from \eqref{kdvgge13} and integrating over $Q_1,Q_3$ in the large $L$ limit, free energy $F=L\, f+o(L)$ is given by 
\be
\label{freef13}
f(\mu_1,\mu_3)=\left.{\cal L} 
\right|_{q_1^*,q_3^*},
\quad {\cal L}=s(q_1,q_3)-\mu_1 q_1-\mu_3 q_3,
\ee
where $q_i^*(\mu_1,\mu_3)$ for $i=1,2$ are defined via saddle point equations 
\bea
\left.{\partial {\cal L}\over \partial q_{2i-1}}\right|_{q_1,q_3^*}=0.
\eea
Using known $1/c$ expansion for $f$, one can then reverse-engineer $s_1$.
This calculation is performed in the Appendix \ref{app:s}.
It is noteworthy that numerically the approximate expression \eqref{naive} is very close to exact result, making corresponding plots visually indistinguishable, see Fig.~\ref{fig:s1}.

To summarize, the  qualitative behavior of $s(\epsilon,c)$ is as follows. We expect function $s(\epsilon)$ for any fixed $c$ to be concave. It reaches its global maximum $s=1$ at $\epsilon$ given by \eqref{epsilonstar}, and in the vicinity of this point 
\be
\label{spert}
s=1-\frac{175 c^2}{8 (4320 c-13552)}\left(\epsilon-\epsilon^*\right)^2+ \mathcal{O}\left(\left|\epsilon - \epsilon^*\right|^3\right)
\ee
This result is derived in the Appendix \ref{app:s}. 
The value at $\epsilon=0$ is also  known exactly, $s(0,c)=\sqrt{1-1/c}$. The behavior inside the interval $0\leq \epsilon \leq 22/5c$ can be parameterized in terms of $1/c$ expansion, 
\bea
s(\epsilon)=1-{s_1(c\epsilon)\over c}+O(1/c^2), 
\eea
with the exact $s_1$ constructed in the Appendix \ref{app:s} and shown in Fig.~\ref{fig:s1}. 
For $\epsilon\sim \mathcal{O}(c^0)$ in the large $c$ limit we expect 
\be\label{eq:holo_entropy}
s(q_1,q_3) = \sqrt{\frac{c\pi q_1}{3}}\left(1-\mathcal{O}(c^{-1/2})\right),
\ee
with the exact behavior going beyond the reach of currently available techniques.

\section{Subsystem entropy for different ensembles}\label{sec:main}
In this section we proceed to employ the semiclassical diagonal approximation discussed in the previous section to compute subsystem entropy of 2d CFT in different global states. We consider CFTs in such states as canonical (Gibbs) and microcanonical ensembles, a primary state, and ensembles carrying KdV  charges: the  KdV Generalized Gibbs Ensemble and KdV generalized microcanonical ensemble.  
We will see that the diagonal approximation reproduces previously known results for the canonical and microcanonical ensembles, and for the primary state. As we discuss below, the latter can be approached as a limit of the KdV (micro)canonical ensemble, resolving apparent contradiction with the Eigenstate Thermalization Hypothesis.

\subsection{Gibbs ensemble}
\label{sec:Gibbs}
For a 2d CFT on an infinite line  in thermal state $\rho\propto e^{-\beta H}$, the subsystem von Neumann and Renyi entropy for one-interval subregion of size $L_A$ were calculated in the famous works of Calabrese 
and  Cardy \cite{Calabrese:2004eu,Calabrese:2009qy} 
\bea
S_A^n={c\over 6}\left(1+{1\over n}\right) \ln \left({\beta\over \pi a}\sinh\left({\pi L_A\over \beta}\right) \right).
\label{Gibbs2d}
\eea
Here $a$ is the UV-cutoff,  and $n$ is the Renyi index, with $n=1$ being the case of von Neumann entropy. 

In the large subsystem limit this result readily follows from \eqref{diagonalap}. We first find the approximation for the reduced Gibbs state
\bea
\label{rhoAGibbs}
\begin{aligned}
    & \rho_A={1\over Z}\int dE\,e^{-\beta E} \int dE_A\;\Omega_{\bar A}(E-E_A)|E_A\rangle \langle E_A| ,\quad
     Z=\int dE\, \Omega(E)\, e^{-\beta E}.
\end{aligned}
\eea
Plugging the Cardy formula \eqref{Omega2d} for the subsystem density of states (\ref{eq:entropy_ansatz}), and evaluating the $E$ integral via saddle-point approximation yields
\bea
\rho_A=\int dE_A \left(e^{-\beta E_A -{c\pi L_A/(12\beta)}}\right) |E_A\rangle \langle E_A|. 
\eea
The Renyi entropy $S_A = (1-n)^{-1}\ln{\text{Tr}\,\rho_A^n}$ can be computed by evaluating the  integral over $E_A$,  also via saddle-point approximation
\bea
\label{RenyiGibbs}
\begin{aligned}
    S_A^n &={1\over 1-n}\ln \int dE_A\, \Omega_A(E_A)\, e^{-n\, \beta E_A - n\, {c\,\pi\, L_A / (12\beta)}} 
    ={c\, \pi L_A \over 12\,\beta}\left(1+{1\over n}\right)+o(L). 
\end{aligned}
\eea
Note, this is only the contribution of the chiral sector. 
For the semiclassical diagonal approximation to be valid, we require $L_A\gg \beta$  \eqref{eq:semiclassical_validity},
which is precisely the limit when \eqref{Gibbs2d} reduces to \eqref{RenyiGibbs}. We emphasize, the agreement with \eqref{Gibbs2d} does not require central charge $c$ to be large.

The canonical ensemble is fully characterized by inverse temperature $\beta$, related to mean energy density as follows
\bea
e\equiv {E\over L}={c\pi\over 12\beta^2}. \label{beta}
\eea
In terms of $\cal S$ defined in \eqref{calS}, after reinstating the anti-chiral sector contribution assuming $\bar c=c$,  we find 
\bea
\label{canonicalS}
{\cal S}(x,n,\beta)={c\pi x \over 6\,\beta}\left(1+{1\over n}\right).
\eea
For the canonical ensemble the semiclassical result \eqref{canonicalS}  is valid for any $0< x < 1$. Our findings are in agreement with the exact results for free field theories  on the torus in the regime $L_A,L\gg \beta$ \cite{Herzog:2013py,Cardy:2014jwa,Datta:2013hba,Chen:2014hta,Chen:2015cna,Chen:2015kua,Chen:2017ahf}.

In the derivation above we ignored the KdV charges. The probability of an individual state $|E\rangle$ in Gibbs ensemble $e^{-\beta H}/Z$  only depends on $E=Q_1$. Therefore Starting from \eqref{diagonalQ} and then integrating over values of higher KdV charges ${Q}_{2i-1}$, $i\geq 2$  would yield the same result, as follows from the discussion in section \ref{sec:semiclassical}. Saddle point values $Q_{2i-1}$ contributing to \eqref{intQmany} with $m=1$ are given by the expectation values of $Q_{2k-1}$ in the canonical ensemble. These were discussed in \cite{Maloney:2018hdg,Dymarsky:2018lhf,Dymarsky:2018iwx}. In the strict thermodynamic limit $L\rightarrow \infty$  one finds for $q_3,q_5$,
\be
\label{onshellq3}
q_3=q_1^2\left(1+{22\over 5c}\right),\; q_5=q_1^3\left(1+{302\over 21c}\right),\; 
q_1={c\pi\over 12\beta^2}.
\ee

Generalization to higher $d$ is straightforward. Repeating (\ref{rhoAGibbs},\ref{RenyiGibbs}) with $\Omega$ given by \eqref{Shd} would readily yield
\bea
S_A^n=\sum_{k=0}^{d-1} {1\over n^k}\,  {\kappa\over d}\,  V_A\, T^{d-1}, 
\eea  
in agreement with the holographic computation of \cite{Dong:2018}.

\subsection{Microcanonical ensemble}\label{sec:ordinary_micro}
For the system in the microcanonical ensemble with the total energy $E$ the approximate reduced density matrix is given by \eqref{diagonalap},
\bea
\rho_A= \int dE_A {\Omega_{\bar A}(E-E_A)\over \Omega(E)}\, |E_A\rangle \langle E_A|.
\eea
Corresponding Renyi entropy can then be obtained by evaluating the $E_A$ integral using saddle-point approximation 
\be
\label{calcmicro}
S_A^n =
{1\over 1-n}\ln \int dE_A\, e^{S(E_A,L_A)+n\, S(E_{\bar A},L_{\bar A})-n\, S(E,L)},
\ee
subject to $E_A+E_{\bar A}=E$. In terms of $\cal S$ defined in \eqref{calS},  after adding anti-chiral sector contribution, we find
\bea
{\cal S}(x,n,\beta)=\left({c\pi \over 3\beta}\right) {\sqrt{x+n^2(1-x)}-n\over 1-n},
\label{micro}
\eea
where $x=L_A/L$, and we introduced effective temperature, as in \eqref{beta}.
The result \eqref{micro} is valid for any $0<x<1$.
As expected for small subsystems $x\ll 1$, function ${\cal S}=\left({c\pi \over 6\beta}\right)\left(1+{1\over n}\right)x+O(x^2)$ is linear, matching Renyi entropy of the canonical ensemble \eqref{canonicalS} due to ensemble equivalence \cite{Dymarsky:2018}.

The saddle-point values of energies $E_A.E_{\bar A}$ that correspond to dominant contribution to  \eqref{micro} are  as follows
\be\label{eq:micro_saddle}
E_A = \frac{x\, E}{x+(1-x)n^2},\quad E_{\bar{A}} =\frac{(1-x)\, n^2\, E}{1+(1-x)n^2}. 
\ee
One can repeat the  calculation of Renyi entropy  for the microcanonical ensemble by taking KdV charges into account and starting from the reduced density matrix \eqref{diagonalQ}. 
Then, first integrating over ${Q}_{2i-1}^A$ and ${Q}_{2i-1}^{\bar A}$ for $i\geq 2$ would readily yield back the integral \eqref{calcmicro}. 
Focusing just on densities of  ${ Q}_1,{ Q}_3$ we thus find 
\bea
q_3^A=(q_1^A)^2\left(1+{22\over 5c}\right),\quad q_3^{\bar A}=(q_1^{\bar A})^2\left(1+{22\over 5c}\right),
\eea
while  $q_1^A\equiv e_A=E_A/L_A$ and $q_1^{\bar A}\equiv e_{\bar A}=E_{\bar A}/L_{\bar A}$ 
are fixed by maximizing $x s(q_1^A)+n (1-x) s(q_1^{\bar A})$ subject to constraint $q_1=q_1^A x+q_1^{\bar A}(1-x)$ with a given 
$q_1\equiv e=E/L={c\pi/(12\beta^2)}$, 
\bea
\label{q1micro}
q_1^A={q_1\over x+n^2(1-x)},\qquad q_1^{\bar A}={q_1 n^2 \over x+n^2(1-x)}.
\eea
From here we find that the states $|E\rangle$ from the microcanonical ensemble contributing the most to $n$-th Renyi entropy have
\be
\label{microq3}
q_3=q_3^A x+q_3^{\bar A}(1-x)= {x+n^4(1-x) \over (x+n^2(1-x))^2} q_1^2 \left(1+{22\over 5c}\right).
\ee
For $n>1$ this is different from \eqref{onshellq3} already at leading order in large $c$ limit.
In this limit  $\epsilon$, defined in \eqref{defepsilon}, for the states described by \eqref{microq3}  is of order one. As is discussed in \cite{Dymarsky:2020,KdVETHholography}, holographically this is the  regime of non-trivial KdV-charged black holes.  In other words states contributing most to Renyi entropy are the descendants of the very high level $n$ scaling with $c$. These are not the statistically most typical states in the (micro)canonical ensemble, which have  $\epsilon \sim 1/c$ and which contribute most to local quantities such as density of energy or density of higher KdV charges, or the von Neumann entropy. This observation emphasizes that higher Renyi entropy is a sensitive probe of the reduced density matrix, which goes beyond conventional local observables. 

In $d>2$ case, Renyi entropy can be obtained in a similar vein using semiclassical diagonal approximation with \eqref{Shd}, 
\be
\label{Renyimicro}
S_A^n={\kappa\, T^{d-1}}V {(x+n^{d}(1-x) )^{1\over d}-n \over (1-n)},\quad x=V_A/V,
\ee
and  $T$ is the effective temperature. The derivation and result \eqref{Renyimicro}  matches the holographic derivation of \cite{Dong:2018}. 

We note that in any $d$ for a small subsystem $x\ll 1$ Renyi entropy for microcanonical and Gibbs ensembles are the same. But for large subsystems  (finite $x$) they are different. The difference disappears for any $x$ in the case of von Neumann entropy $n\rightarrow 1$, as was emphasized in \cite{Grover:2018,Dymarsky:2018}.

\subsection{Primary eigenstate ``ensemble''}\label{sec:primary_ensemble}
For chaotic CFTs in $d>2$, the cases of microcanonical and eigenstate ensembles are similar, in both cases
when $x\leq 1/2$, within the diagonally-dominated semiclassical approximation  one obtains the same expression for the subsystem Renyi entropy \eqref{Renyimicro}. This matches holographic calculation of Renyi entropy for $d>2$ CFT in an eigenstate \cite{Dong:2020iod,Dong:2023bfy}, which used the ``fixed area states'' approximation -- the holographic analog of the semiclassical diagonal approximation \eqref{diagonalap}.

The $d=2$ case is different due to additional conservation laws from the KdV charges. In particular, starting with  a  heavy primary CFT state $|\Delta\rangle$ with fixed energy density  $e=2\pi \Delta/L^2$, $L\rightarrow \infty$, we do not expect its decomposition \eqref{eq:decomposition_KdV}, assuming it can be rigorously defined,  to include all possible subsystem states, but only very special ``primary'' ones. This can be understood as follows. In addition to being an energy eigenstate, the primary states are simultaneous eigenstates of all KdV charges, saturating inequalities \eqref{qineq} at the level of charge  densities 
\be \label{eq:KdV_inequality}
q_{2k-1}=q_1^k.
\ee 
We remark satisfying (\ref{eq:KdV_inequality}) for any $k\geq 1$ would  automatically imply it holds for all $k$.
These conditions are satisfied locally, and therefor 
for large subsystems, assuming subsystem eigenstates can be defined, decomposition \eqref{eq:decomposition_KdV} would only include subsystem  states 
$|\vec{Q}_A\rangle , |\vec{Q}_{\bar A}\rangle$ satisfying 
\be\label{eq:KdV_inequality_sub}
q^A_{2k-1} =\left(q^A_1\right)^k,\;\; q^{\bar{A}}_{2k-1} = \left(q^{\bar{A}}_1\right)^k.
\ee  
These are essentially primary states of the subsystem Hamiltonians, which we denote by $|\Delta_A\rangle, |\Delta_{\bar A}\rangle$, 
with $q^A_1=\Delta_A/L_A^2$ and similarly for $\bar A$.
In other words, the primary states decomposed into subsystem states are expected to take the special form 
\bea 
|\Delta\rangle &=& \sum_{\vec{q}_A,\vec{q}_{\bar{A}}}C^\Delta_{\Delta_A\;\Delta_{\bar{A}}} \times  |\Delta_A\rangle \otimes |\Delta_{\bar{A}}\rangle, 
\eea
where coefficients $C^\Delta_{\Delta_A\;\Delta_{\bar{A}}}$ enforce equality of all KdV charge densities
\be\label{eq:subsystem_primary}
q^A_1 = q^{\bar{A}}_1 = q_1\equiv \Delta/L^2,\;\;q^{A,\bar{A}}_{2k-1} = \left(q^{A,\bar{A}}_1\right)^k,\;\;k\geq 1,
\ee 
but otherwise are assumed to be structureless. 
As a result, the reduced density matrix is essentially the subsystem ``primary microcanonical'' ensemble, with only subsystem primary states from a narrow energy window contributing with equal weights,
\bea
\begin{aligned}
    \rho_A(\Delta) &= {\rm Tr}_{\bar A}(|\Delta \rangle \langle \Delta |) ={1\over Z}\int d \Delta_A \Omega_p(\Delta_A) |\Delta_A\rangle\langle \Delta_A|\label{mcA},\quad
    Z = \int d \Delta_A \Omega_p(\Delta_A).
\end{aligned}
\eea 
Here the integrals are over a narrow window of states satisfying $\Delta_A/L_A^2=\Delta/L^2$ and $\Omega_p$ is the Cardy formula for the density of primary states, c.f.~\eqref{eq:Cardy_corrected}, 
\bea
\ln \Omega_p=\pi \sqrt{2(c-1)\Delta/3}.
\eea
From here, by repeating the steps described in Section \ref{sec:semiclassical}, we readily find that Renyi entropy is index independent, 
\bea
\label{primaryRenyi}
S^n_A(\Delta)&=&L_A \sqrt{(c-1)\pi e\over 3},\qquad e={2\pi\Delta\over L^2}.
\eea
By introducing effective temperature \eqref{beta} associated with the energy density $e$ and adding contribution  of the anti-chiral sector 
we can rewrite this as 
\bea
\label{primaryRenyibeta}
S^n_A(\Delta)&=&L\, {\cal S},\quad {\cal S}(x,n,\beta)={\sqrt{c(c-1)}\pi \over 3\beta}x. 
\eea
For a primary state in a chaotic CFT this result is valid for $x\leq 1/2$ with the entropy for $x>1/2$ given by substituting $x\rightarrow 1-x$. Starting from the ``microcanonical primary'' ensemble instead, a superposition of  primary states within a narrow energy window with equal coefficients, reduced entropy will be given by \eqref{primaryRenyibeta} for any $0<x<1$.

It is interesting to note that  \eqref{primaryRenyibeta} agrees with the von Neumann entropy for thermal state \eqref{RenyiGibbs} at the leading order in $c$. From a holographic perspective, the match between  von Neumann entropy of primary and thermal states is expected to hold for any $L_A$ due to the fact that in absence of the cosmic brane back-reaction bulk geometry in both cases is the same, i.e.~the BTZ black hole. Agreement at leading order in $1/c$ continues to the regime of small $L_A/\beta$ as was verified in \cite{He:2017}.

The most notable feature of (\ref{primaryRenyi}) is the index independence of $S^n_A$. For holographic CFTs at the leading order in $c\to \infty$, there exists an alternative computation of $S^n_A(\Delta)$ based on the monodromy method, by computing a correlation function 
involving twist operators,
\be
S^n_A(\Delta) = \frac{1}{1-n}\ln{\text{Tr}\rho^n_A(\Delta)},\quad \text{Tr} \rho^n_A(\Delta) \propto \langle \mathcal{O}_\Delta \sigma_n \sigma_n \mathcal{O}_\Delta \rangle. 
\ee 
In addition to large $c$, one should still  take high energy limit $\Delta \gg 1$ and large subsystem size. This computation was performed in \cite{Wang:2018}, and it is technically independent of the diagonally-dominated semiclassical approximation we have used above. The results of \cite{Wang:2018} is in full agreement with the large central charge limit of (\ref{primaryRenyi}). In the context of AdS/CFT,  $n$-independence  of  $S^n_A$ suggests that the dual bulk  configurations are, at leading order, the fixed-area states \cite{fixedarea1,fixedarea2,fixedarea3,TN1,TN2}.  Such states provide a useful basis to understand the phase space structure of the semiclassical gravity in the bulk. In other words, in the case of 2d CFTs heavy primary states provide  an explicit example of fixed area holographic  states. 

The discrepancy between the subsystem Renyi entropy for $n>1$ in the microcanonical  ensemble and in primary eigenstate already at the leading order in large $c$ is a stark violation of the Eigenstate Thermalization Hypothesis in 2d CFT. The reason for the discrepancy is as follows. The states contributing most to the subsystem Renyi entropy in the microcanonical case have $q_3$  given by (\ref{microq3}). 
This is strictly greater than $q_3=q_1^2$ of primary states. The states obeying \eqref{microq3} are descendants of the very high level $n \sim L^2q_1$. They have very different physical properties from the primary states. 
As was suggested in \cite{Lashkari:2017hwq,He:2017txy,Dymarsky:2018lhf}  including KdV charges into account should resolve this discrepancy by matching the primary states to KdV-generalized canonical or microcanonical ensembles. This, too, comes with a subteltly. We discuss this below in sections \ref{sec:KdVGGE} and \ref{sec:KdV-micro}.

\subsection{KdV Generalized Gibbs Ensemble}
\label{sec:KdVGGE}
Before turning to KdV microcanonical ensembles, we consider subsystem Renyi entropy for CFTs in KdV generalized Gibbs ensemble (KdV GGE), i.e. we consider the state of the form
\bea
\label{kdvgge}
\rho = e^{-\sum \mu_{2i-1}{Q}_{2i-1}-F(\mu)}, \quad {\rm Tr}\,\rho=1,
\eea
where $\mu_{2i-1}$ label the set of KdV chemical potentials. The semiclassical diagonal approximation (\ref{diagonalQ}) for evaluating subsystem entropy would require $L^{2k-1}_A \gg \mu_{2k-1}$. This is automatically satisfied provided chemical potentials are fixed, together with $x=L_A/L$,  while $L\rightarrow \infty$.
In this regime, the extensive part of the subsystem Renyi entropy can be expressed in terms of the density of generalized free energy of the full system $F=Lf(\mu_{2k-1})$  as follows
\bea
\label{SGGE}
S_A^n =L_A{f(n\mu_{2k-1})-nf(\mu_{2k-1})\over 1-n}.
\eea
This is of course a completely universal result for any system with extensive conserved quantities in the corresponding GGE state. 

For 2d CFT free energy density $f$ of KdV GGE  was evaluated in \cite{Dymarsky:2018iwx} in large $c$ limit, including first non-trivial $1/c$ correction, 
\bea
\label{freeE}
f={c\, \pi \over 12 \mu_1}(f_0+f_1/c+\dots),
\eea
where $f_0,f_1,...$ each admit a perturbative expansion in $t_{2k-1}$ defined by
\be 
\label{tdef}
t_{2k-1}=\left(\frac{c\pi}{12}\right)^{k-1} \frac{\mu_{2k-1}}{\mu_1^{2k-1}}. 
\ee
In general case the  computation of \cite{Dymarsky:2018iwx} is valid only as an expansion for small $t_{2k-1}$. The leading order term $f_0$ receives the contribution from the states satisfying $q_{2k-1}=q_1^k$. In holography, this corresponds to free energy of a BTZ black hole. The sub-leading term $f_1$ then encodes one-loop graviton corrections on top of the black hole background. From the CFT perspective, these states are the ``near-primary'' states with $\epsilon$ defined in \eqref{defepsilon} of order $1/c$ -- the same scaling with $c$ as the most typical states in the ordinary microcanonical ensemble. 
When $t_{2i-1}$ are not very small,  for generic combinations of the KdV chemical potentials $\mu_{2k-1}$, the BTZ configurations are not always dominant saddles of the Euclidean gravity path integral dual to KdV GGE, and there are explicit examples when the leading contribution to $f$ is given by a non-trivial KdV-charged black hole configuration \cite{Dymarsky:2020}. Their roles in computing the holographic Renyi entropy will be discussed in more details in the companion paper \cite{KdVETHholography}. 

To illustrate various regimes of index dependence of $S_A^n$ we focus on the KdV GGE with only two chemical potentials $\mu_1,\mu_3$ 
turned on. In this case there are no classical KdV-charged black hole saddled, except for the BTZ ones \cite{Dymarsky:2020} and both $f_0,f_1$ can be found in a closed form, see Appendix \ref{app:f}, as a function of real-valued
\bea
\label{tau}
\tau={\mu_1\over \mu_3^{1/3}}\left(12\over c\pi\right)^{1/3}.
\eea
The KdV GGE is well defined for any positive $\mu_3$ and arbitrary $\mu_1$, such that $\tau$ can take any real value. When $\tau>0$, it is related to \eqref{tdef} via $t_3=1/\tau^3$. Explicit form of free energy\footnote{Although we use the same notations $f_i$ as in \eqref{freeE}, these are different functions, related by change of variables, $f_i(t_3)=\tau f_i(\tau)$.}  
\bea
\label{ftau}
f=\left({c\pi \over 12 }\right)^{2/3} \mu_3^{-1/3}(f_0(\tau)+f_1(\tau)/c+O(1/c^2))
\eea
is used in the Appendix \ref{app:s} to fix leading  $1/c$ order  of the KdV microcanonical entropy \eqref{scexp}.

Before we discuss different regimes, we note that von Neumann entropy $n=1$ of the KdV GGE ensemble at leading order in $1/c$ is the same as von Neumann entropy of the conventional thermal ensemble. Technically, this follows from the properties of function $f_0$. A simple alternative way to see that is to combine \eqref{freef13} with \eqref{SGGE} in the limit $n \rightarrow 1$, which yields
\bea
S_A^{vN}=L_A\, s(q_1,q_3). 
\eea
Equality with thermal entropy follows because at leading $1/c$ order $s(q_1,q_3)$ is only $q_1$-dependent, given by Cardy formula, see \eqref{eq:holo_entropy}.  

To probe various regimes of index dependence, we first consider the case of small $\mu_3$, i.e.~a  small perturbation by $Q_3$ of the conventional CFT Hamiltonian. In this regime \cite{Dymarsky:2018lhf},
\be
\begin{aligned}
     f_0=1-t_3+4 t_3^2+O(t_3^3), \quad
     f_1=-{22\over 5}t_3+{2096\over 35}t_3^2+O(t_3^2),
\end{aligned}
\ee
and for  $i\geq 2$, $f_i=O(t_3^i)$.  From $q_{2i-1}=-\partial f/\partial \mu_{2i-1}$ we readily find effective temperature $\beta$ and parameter $\epsilon$ \eqref{defepsilon}, 
\bea
\beta &=& \mu_1 \left(1+2\left(1+{22\over 5c}\right)t_3+O(t_3^2)\right),\\
c \epsilon &=& {22\over 5}-\left({1728\over 35}+O(1/c)\right)t_3+O(t_3^2).
\eea
In terms of effective temperature and $t_3$, $n$-th Renyi entropy is (both chiral and anti-chiral sectors are included) 
\be
\begin{aligned}
     {\cal S}(x,n,\beta, t_3) = x\left({c\pi\over 6\beta}\right)\Bigg( 1+{1\over n} - 
     \left(1+{22\over 5c}\right){(n-1)(1+n)^2\over n^3}t_3 +O(t_3^2)\Bigg).
\end{aligned}
\ee

Next we consider the case of $\mu_1=0$ or $\tau=0$ defined in \eqref{tau}. In terms of effective temperature $q_1(\mu_3)={c\pi\over 12\beta^2}$ we find Renyi entropy, see Appendix \ref{app:f} for details,
\bea
\label{st0}
\begin{aligned}
    {\cal S}(x,n,\beta) =x \left({c\pi\over 4\beta}\right) \left(1-\frac{0.225153}{c}+O(1/c^2)\right)
     {1+n^{1/3}+n^{2/3}+n\over n^{1/3}(1+n^{1/3}+n^{2/3})}.
\end{aligned}
\eea

As we see from \eqref{SGGE} and the examples above, in general  the subsystem Renyi entropy  for a theory in the KdV GGE is explicitly index-dependent. This is  in a stark contrast with the result for  the primary eigenstate in section \ref{sec:primary_ensemble}. 
The discrepancy seemingly suggests that KdV-generalized ETH breaks down: the latter would require local properties of eigenstates, including primary states, to be described by a certain GGE. This issue was observed and resolved  in  the original work \cite{Dymarsky:2018lhf}, which pointed out that the primary states should correspond to singular GGE with at least some chemical potentials approaching infinity. Consider a GGE with only two chemical potentials, $\mu_3$ and $\mu_1$,   in the limit 
$\mu_1\rightarrow -\infty$ with $|\mu_1|/\mu_3$ fixed. 
As is discussed in \cite{Dymarsky:2019etq} this GGE will describe primary states with $q_1={-\mu_1/(2\mu_3)}(1+o(1/c^2))$ and $q_3=q_1^2$, see Appendix \ref{app:f}. In this limit the GGE free energy density is given by 
\be
\label{primaryf}
f={\mu_1^2\over 4\mu_3}+\left({c\pi^2\over 6}\right)^{1/2}\sqrt{2|\mu_1|\over \mu_3}\left(1-{1\over 2c}+O(1/c^2)\right).
\ee
It follows from \eqref{SGGE} that the first term in \eqref{primaryf} does not contribute to Renyi entropy, while second term gives index-independent contribution
\bea
{\cal S}(x,n,\beta)=x \left({c\pi\over 3\beta}\right)\left(1-{1\over 2c}+O(1/c^2)\right), 
\eea
in full agreement with \eqref{primaryRenyibeta}.

\subsection{KdV microcanonical  ensemble} 
\label{sec:KdV-micro}
In what follows we focus on the generalized KdV-microcanonical ensemble with only first two KdV charges ${Q}_1,{Q}_3$ fixed. 
The procedure to evaluate subsystem Renyi entropy is conceptually straightforward. In the diagonally-dominated semiclassical approximation, the microcanonical Renyi entropy is given by the integral 
\be
\begin{aligned}
     & S^n_A(q_1,q_3) =\frac{1}{1-n}\ln \bigg(Z^{-n}\int \left[\prod dQ^{A}_{1,3}\right]\; 
    \; e^{Lxs(q^A_1,q^A_3)+nL(1-x)s(q^{\bar{A}}_1,q^{\bar{A}}_3)}\bigg),\\
    & \qquad Z=e^{Ls(q_1,q_3)},\qquad q_{2i-1}=Q_{2i-1}/L.
\end{aligned}
\ee
The integral is over two variables $q^{A}_{1,3}$, while  $q^{\bar A}_{1,3}$ are fixed by the conservation constraints
\bea\label{eq:conservation1}
q_1&=&q_1^A x+q_1^{\bar A} (1-x),\\
q_3&=&q_3^A x+q_3^{\bar A} (1-x).
\label{eq:conservation3}
\eea
In the thermodynamic limit $L\rightarrow \infty$ the saddle point is given by minimizing the ``effective action'' (factor $2$ to include both chiral and anti-chiral sectors) 
\be
\label{effa}
{\cal S}=2\left[{x\, s(q^A_1,q^A_3)+n(1-x)\, s(q^{\bar{A}}_1, q^{\bar{A}}_3) -n\, s(q_1,q_3)\over 1-n}\right]
\ee
with respect to $q^A_{1,3}$ and $q^{\bar A}_{1,3}$ with  given fixed $q_{1,3}$ and subject to constraints (\ref{eq:conservation1},\ref{eq:conservation3}).

Before we discuss the general case, we consider the holographic limit of large $c$. As we discussed in section \ref{SQ1Q3}, working in the regime when $q_{2k-1}\propto c^k$, at leading $1/c$ order entropy $s(q_1,q_3)$ is $q_3$-independent, see \eqref{eq:holo_entropy}. Writing it as a function of $q_1$ and $\epsilon$ as in \eqref{sq1q3}, $s(\epsilon,c)$ is a monotonically decreasing function of $\epsilon\geq 0$ reaching its maximum at $s(0)=1$.

Since  \eqref{effa} depends on $q_3^A,q_3^{\bar A}$ only at the subleading order in $1/\sqrt{c}$,
we can first ignore $q_3^A,q_3^{\bar A}$ as well as  the constraint in \eqref{eq:conservation3} and maximize $\cal S$ with respect to $q_1^A,q_1^{\bar A}$ subject only to $q_1=q_1^A x+q_1^{\bar A} (1-x)$ using 
$s(q_1,q_3)=\sqrt{c\pi q_1/3}$. This is exactly the calculation of section \ref{sec:ordinary_micro} for the microcanonical ensemble, and the resulting values of $q_1^A,q_1^{\bar A}$ are given by \eqref{q1micro}. 

Going back to the constraint \eqref{eq:conservation3}, together with the inequality \eqref{qineq} it would imply $q_3\geq q_3^{\rm crit}$, where, c.f.~with \eqref{microq3}, 
\bea
q_3^{\rm crit}(x,n,q_1)= {x+n^4(1-x) \over (x+n^2(1-x))^2} q_1^2.
\eea
This is the condition which should be satisfied for self-consistency.
In other words, for the given densities  of the global state $q_1,q_3$, 
provided $q_3> q_3^{\rm crit}(x,n,q_1)$,  maximization at leading order in $c$ will fix 
$q_1^A,q_1^{\bar A}$  to the same values as in the case of the microcanonical ensemble \eqref{q1micro}, 
while $q_3^A,q_3^{\bar A}$ will be given by\footnote{The following expressions are derived using the explicit  form of the entropy $s(\epsilon)$ given by \eqref{slargec}. The latter can not be rigorously justified at this point. More generally, assuming only concave property of $s(\epsilon)$ in the $c\rightarrow \infty$ limit one can  show $q_3^A,q_3^{\bar A}$ maximizing $\cal S$ will satisfy $q_3^A\geq (q_1^A)^2,q_3^{\bar A}\geq (q_1^{\bar A})^2$ and will be unique. At leading order the on-shell value of $\cal S$ remains the same.} 
\bea
q_3^A-(q_1^A)^2=q_3^{\bar A}-(q_1^{\bar A})^2 =q_3-q_3^{\rm crit}.
\eea
Since $\cal S$ depends on $q_3$ only at the subleading order $1/\sqrt{c}$, 
resulting Renyi entropy, at leading in $c$ order, will be the same as in the case of micro-canonical ensemble \eqref{micro},
\bea
S_A^n=L\sqrt{c\pi q_1\over 3} \left({\sqrt{x+n^2(1-x)}-n\over 1-n}\right).
\label{microkdv}
\eea

When $q_1^2\leq q_3\leq  q_3^{\rm crit}(x,n,q_1)$ the inequality \eqref{qineq} at the level of each subsystem, $A$ and $\bar A$, will force $q^A_3=(q^A_1)^2$, 
$q^{\bar A}_3=(q^{\bar A}_1)^2$,  fixing $q_1^A,q_1^{\bar A}$ from the constraints (\ref{eq:conservation1},
\ref{eq:conservation3}) 
\be\label{eq:glued_BTZ}
q^A_1 = q_1 \left(1- \sqrt{\frac{(1-x)\,\epsilon}{x}}\right),\;\;q^{\bar{A}}_1 = q_1 \left(1+\sqrt{\frac{x\,\epsilon}{1-x}}\right), 
\ee
and $\epsilon$ is defined by \eqref{defepsilon}. Notice that so far $\epsilon$ is small enough such that 
$q_3< q_3^{\rm crit}(x,n,q_1)$, the expression for $q_1^A$ is positive, as necessary for self-consistency.  
There is another saddle point, related by a sign flip in \eqref{eq:glued_BTZ}; it yields a subleading contribution to entropy. More specifically, it can be checked that for $q_1^2\leq q_3\leq q^{crit}_3$, second branch always gives rise  to a larger value of the effective action (\ref{effa}). 

It is interesting to note, that in contrast to \eqref{q1micro}, 
subsystem  densities (\ref{eq:glued_BTZ}) are $n$-independent. In AdS/CFT, we can interpret these saddle-point configuration as gluing two different BTZ black hole segments, one for subsystems $A$ and $\bar{A}$.  For this reason we will refer to \eqref{eq:glued_BTZ} as the ``glued-BTZ'' solutions. More details about the bulk picture can be found in the companion paper \cite{KdVETHholography}.

For the glued-BTZ configuration the value of Renyi entropy at the leading order in large $c$ is 
\be\label{eq:Renyi_BTZ}
S^n_A =  L \sqrt{\frac{c\pi}{3}}\left[\frac{x\sqrt{q^A_1}+n(1-x)\sqrt{q^{\bar{A}}_1}-n \sqrt{q_1}}{1-n}\right].  
\ee
We note, that even though (\ref{eq:Renyi_BTZ}) is explicitly $n$-dependent, corresponding refined Renyi entropy \cite{Dong:2016renyi}  defined by
\be 
\tilde{S}^n_A = n^2\partial_n \left(\frac{n-1}{n}S^n_A\right) = x  L \sqrt{\frac{c \pi q^A_1}{3}}
\ee
is $n$-independent. Holographically, it indicates that the  glued-BTZ solution in the bulk remains the same for different values of the cosmic-brane tension, which backreacts on the geometry. The backreaction independence is a remarkable property also shared by fixed-area states \cite{fixedarea1,fixedarea2,fixedarea3}.

Combining all together, we  find that in the holographic limit  the  subsystem Renyi entropy of the KdV microcanonical ensemble exhibits a sharp transition between the ordinary microcanonical value (\ref{Renyimicro}) and that of the glued-BTZ solutions (\ref{eq:Renyi_BTZ}) 
\begin{numcases}{{\cal S}(x,n,\beta,\epsilon) = \left({c\pi\over 3\beta}\right)}
 \resizebox{0.17\textwidth}{!}{$\frac{\sqrt{x+n^2(1-x)}-n}{1-n}$,}
       \qquad\qquad\qquad\qquad\qquad\qquad\quad ~\,\epsilon \geq \epsilon_{crit}, \label{eq:KdV_Renyi_holo1}\\
\resizebox{0.5\textwidth}{!}{$\frac{\sqrt{x\left[x-\sqrt{x(1-x)\epsilon}\right]}+n\sqrt{(1-x)\left[(1-x)+\sqrt{x(1-x)\epsilon}\right]}-n}{1-n}$},
     ~~  \epsilon \leq \epsilon_{crit}, \label{eq:KdV_Renyi_holo2}
\end{numcases}
where the densities of the original state $q_1,q_3$ are written in terms of the effective temperature $\beta$ and $\epsilon=q_3/q_1^2-1$, and where
\be
\label{epsiloncrit}
\epsilon_{crit}(x,n)\equiv {q_3^{crit}(x,n,q_1)-q_1^2\over q_1^2}={(n^2-1)^2x(1-x)\over (x+n^2(1-x))^2}.
\ee
This result is valid for $0<x<1$. For chaotic CFTs obeying KdV ETH, it should also be applicable for $x<1/2$  to ``typical" eigenstates characterized by densities $q_1,q_3$, whose higher Kdv charge densities  satisfy: $q_{2i-1}=q^*_{2i-1}$ for $i>m=2$ from \eqref{sps}.

The transition between \eqref{eq:KdV_Renyi_holo1} and \eqref{eq:KdV_Renyi_holo2} can be understood as a  function of $\epsilon$, $x$ or $n$. Say, for a given global state with  fixed $\beta$ and $\epsilon$, and given subsystems $A$, $\bar A$ such that $x$ is fixed, we can discuss index dependence of Renyi entropy. 
When $\epsilon \geq {x\over 1-x}$ there is no transition, and the microcanonical result \eqref{microkdv} applies for all $n$. When 
$\epsilon < {x\over 1-x}$, there is a critical value of index $n_{crit}$ defined by 
\bea
\epsilon=\epsilon_{crit}(x,n_{crit}), \label{eq:n_crit}
\eea
such that \eqref{microkdv} applies for $n\leq n_{crit}$ while  for $n\geq n_{crit}$ the entropy is given by  \eqref{eq:Renyi_BTZ}. 
This matches   holographic computation of the companion paper \cite{KdVETHholography}, which finds a transition between the on-shell bulk configurations happening at  the critical value
\be 
n_{crit} = \sqrt{q^{\bar{A}}_1/q^A_1},
\ee
which can be checked to agree with (\ref{eq:n_crit}). For $n>n_{crit}$, relevant bulk configuration is the glued-BTZ geometry. As $n$ decreases, the backreaction of the cosmic brane becomes smaller and at  $n=n_{crit}$ the glued-BTZ configuration becomes unstable. Classical saddles for $n<n_{crit}$ are not known explicitly \cite{KdVETHholography}.
\begin{figure}
    \centering
    \includegraphics[width=0.85\textwidth]{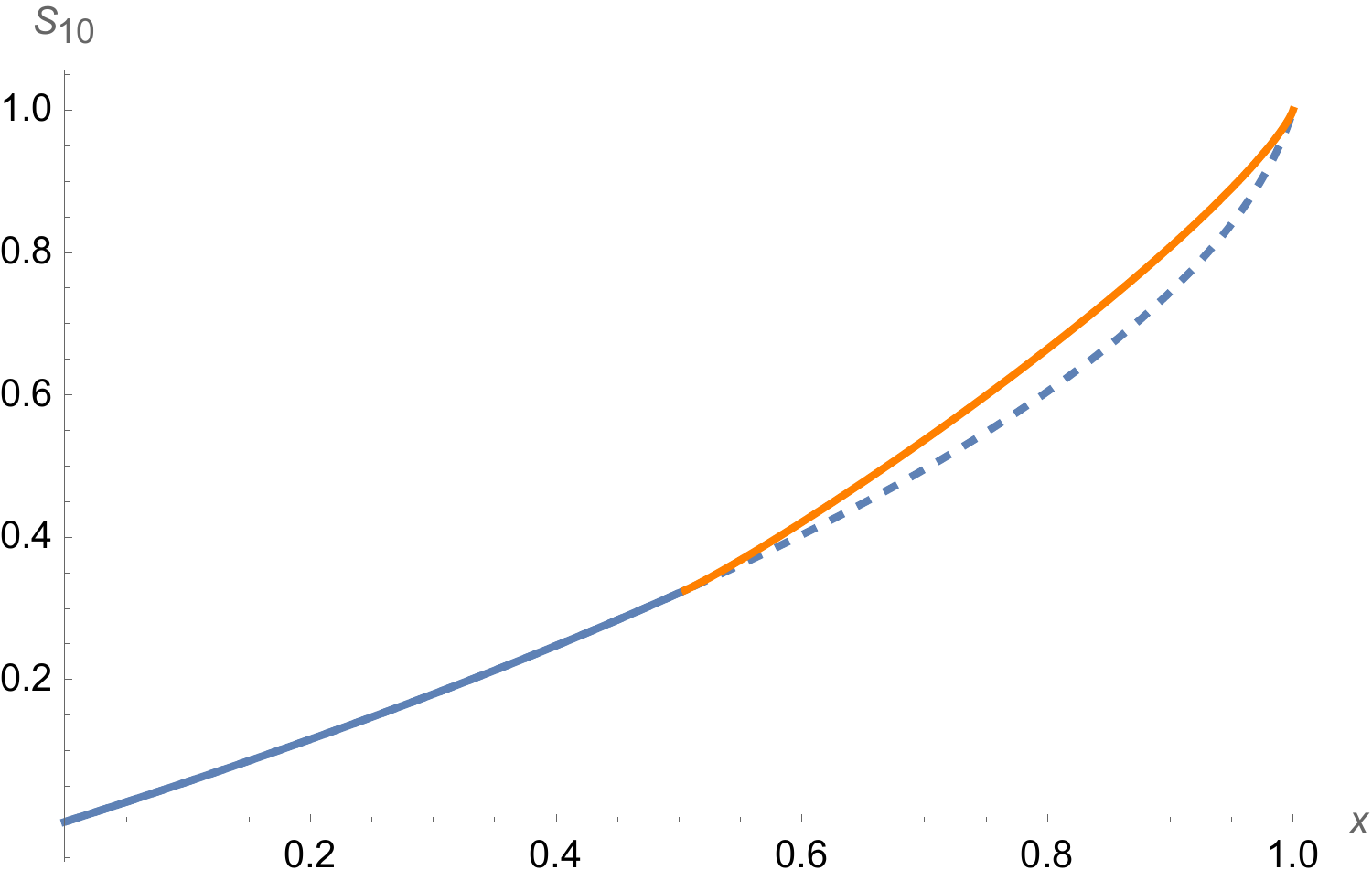}
    \caption{Tenth Renyi entropy ${\cal S}(x,n=10,\beta,\epsilon)/(c\pi^2/3\beta)$ given by (\ref{eq:KdV_Renyi_holo1}-\ref{eq:KdV_Renyi_holo2}), as a function of $x$ for $\epsilon=392/400$. For $x_l\leq x\leq x_r$, where $x_l=50/99$, $x_r=9800/9801$, Renyi entropy is given by \eqref{eq:KdV_Renyi_holo2} (shown in orange) while for all other values \eqref{eq:KdV_Renyi_holo1} applies (shown in solid blue). Values of \eqref{eq:KdV_Renyi_holo1} inside the interval $x_l\leq x\leq x_r$  (dashed blue) are shown for comparison.}
    \label{fig:S2}
\end{figure}

When $\epsilon \rightarrow 0$, the critical value of index $n_{crit}\rightarrow 1$ 
such that Renyi entropy for any $n\geq 1$ is given by \eqref{eq:KdV_Renyi_holo2}, which becomes $n$-independent, matching \eqref{primaryRenyibeta} at leading order in $c$. This result is more general and holds for finite $c$. In the $\epsilon \rightarrow 0$ limit the KdV-microcanonical ensemble reproduces the result for the  primary state see Appendix \ref{app:finite_c}, resolving an apparent tension between the results of \cite{Wang:2018} and the KdV ETH.  

It is also instructive to fix $\epsilon$ and $n$ and discuss the shape of  the Page curve, i.e.~the $x$-dependence of $\cal S$. When 
\bea
\epsilon \geq {(n^2-1)^2\over 4n^2},
\eea
the entire Page curve is given by \eqref{eq:KdV_Renyi_holo1}, matching the microcanonical ensemble result. In the opposite case, there are two values $x_{l,r}$, $x_l<x_r$, satisfying $\epsilon=\epsilon_{crit}(x_{l,r},n)$, such that the microcanonical result applies for $0\leq x\leq x_{l}$ and $x_r\leq x\leq 1$, while  \eqref{eq:KdV_Renyi_holo2} applies for $x_l\leq x\leq x_r$.
We show this behavior in Fig.~\ref{fig:S2}, where we plot $n=10$-th Renyi entropy for the KdV-microcanonical ensemble with $\epsilon=392/400$.

At finite $c$ qualitative behavior of the subsystem Renyi entropy remains the same as in the large $c$ limit discussed above. The crucial difference, when $c$ is large but finite, sharp transition between the glued-BTZ and the microcanonical results becomes a crossover around $n\approx n_{crit}$ of size $\Delta n\sim 1/\sqrt{c}$. 
As we show in the Appendix \ref{app:finite_c} using qualitative behavior of $s(q_1,q_3)$,  for any $c$ when $n\rightarrow 1$, $S^{n}_A$  approaches the micro-canonical von Neumann entropy $S^{vN}_A$. For $n\gg n_{crit}$,  $S^{n}_A$ as a function of $n$ admits an expansion in  $1/n$,  $S^n_A=S^\infty_A-O(1/n)$, see the Appendix \ref{app:KdVmicro}.
When $\epsilon$ decreases, the values of $S^n_A$ become bounded within a narrow window of width  $\Delta S_A \propto \sqrt{\epsilon}\, S^{vN}_A$, where $\Delta S_A = S^{vN}_A-S^{\infty}_A$. 
Also, the critical  value of $n_{crit}$, marking the crossover region, decreases towards $n=1$, $n_{crit}-1\propto \sqrt{\epsilon}$. We illustrate this behavior numerically in the Appendix \ref{app:KdVmicro}, where we use the approximate expression for the density of states (\ref{eq:KdV_entropy}) to evaluate $S_A^n$ as a function of $n$ for various $c$. By increasing $c$ numerically, we show how a smooth crossover region shrinks and  a sharp transition near $n=n_{crit}$ develops.

\begin{figure}[ht]
    \centering
    \includegraphics[width=0.49\linewidth]{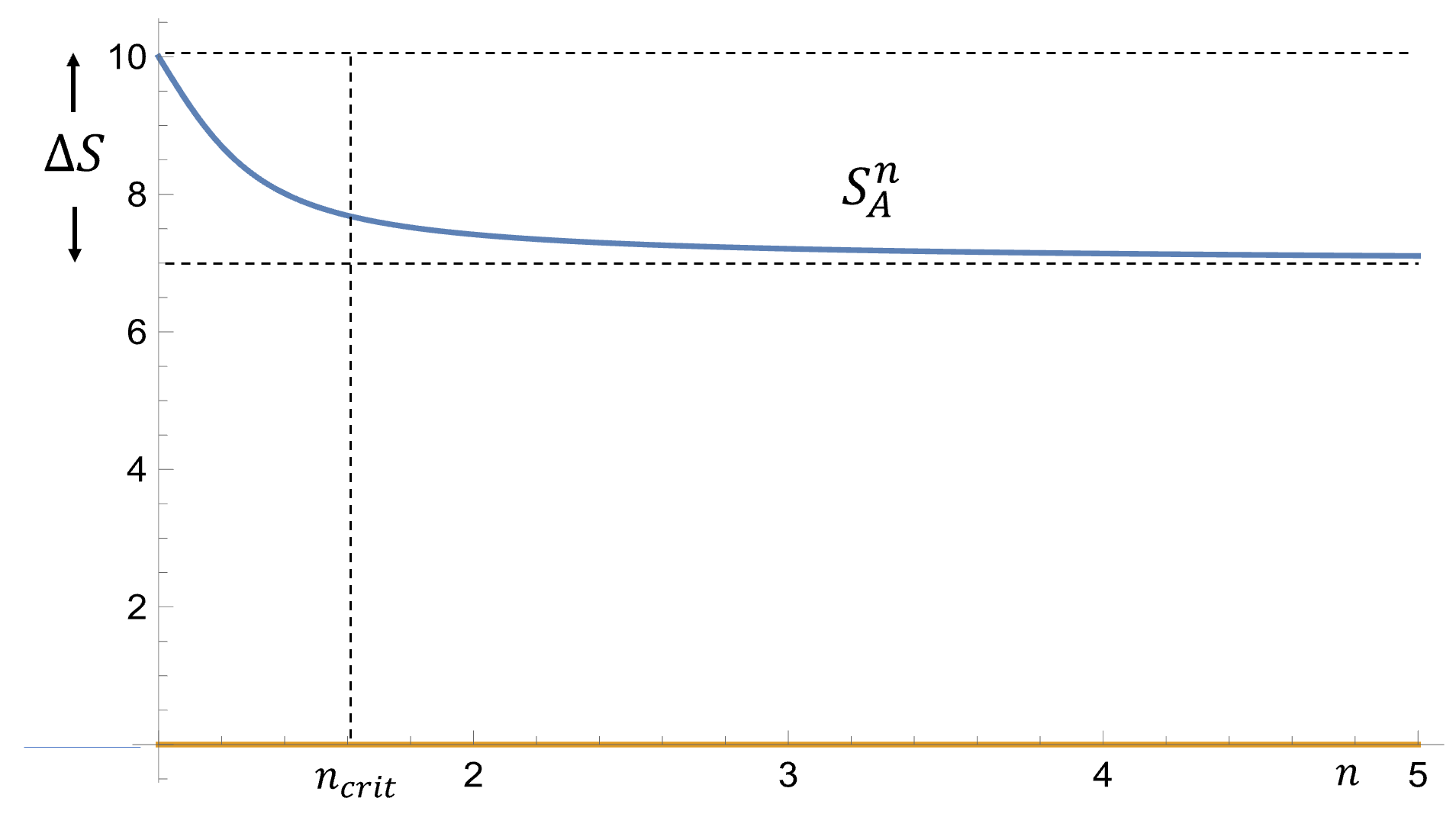}
    \includegraphics[width=0.49\linewidth]{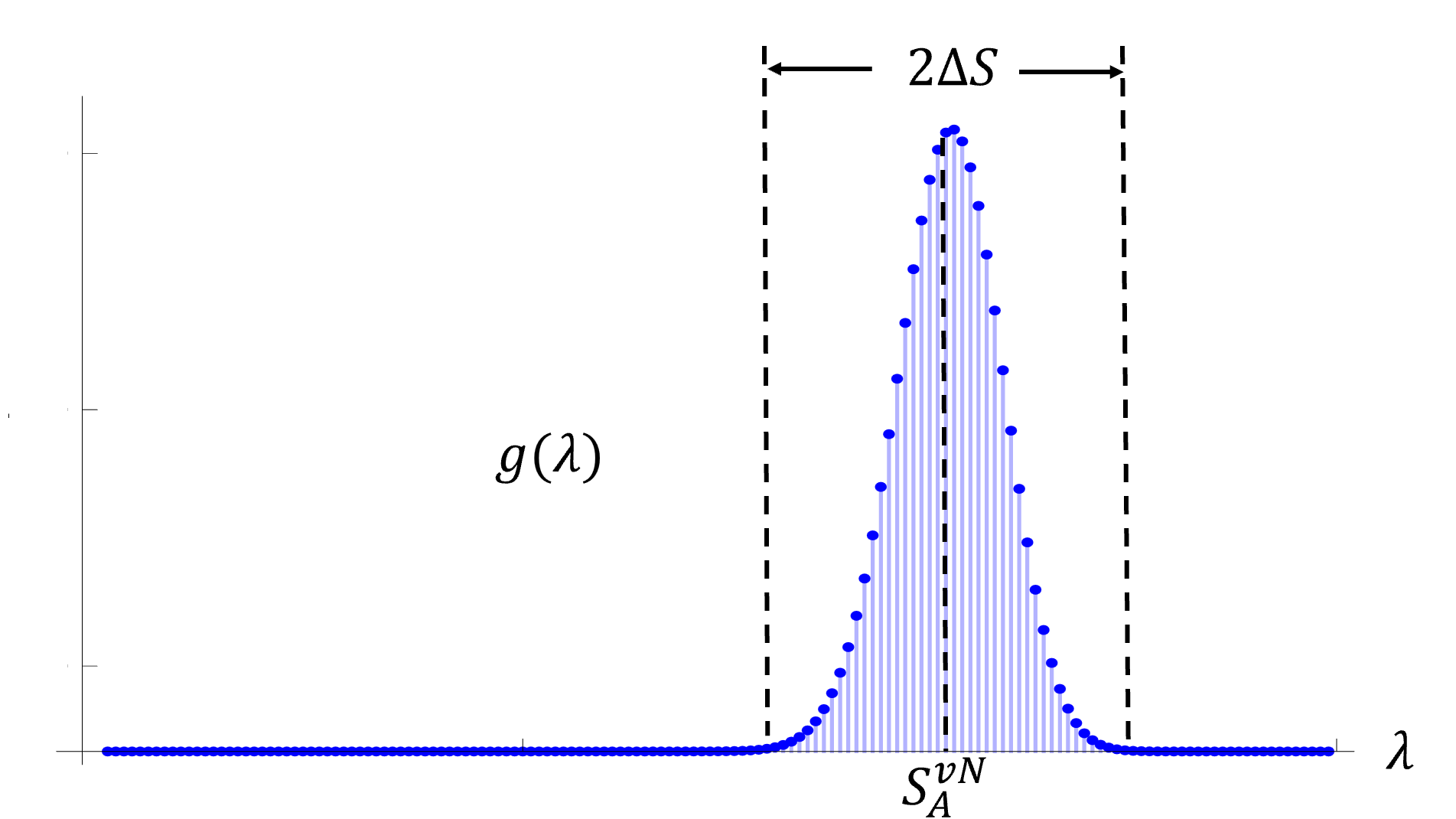}
    \caption{Left: qualitative behavior of  the subsystem Renyi entropy, with $n$-dependence being  polynomially suppressed for  $n\gg n_{crit}$,  $S^n_A=S_A^\infty -O(1/n)$. 
        Right: qualitative behavior of the corresponding entanglement spectral density $g(\lambda)$, which is sharply peaked around $S_A^{vN}$.}
    \label{fig:renyi_to_spectrum1}
\end{figure}

The suppression of index dependence of $S^{n}_A$ in the limit of small $\epsilon\ll 1$, when $\Delta S_A/S_A^{vN} \ll 1$,  has a clear interpretation in terms of the entanglement spectrum  of the reduced density matrix
\be
\rho_A = \sum_{\lambda} e^{-\lambda} |\lambda \rangle \langle \lambda |.
\ee
In a continuous limit  when the sum over $\lambda$ becomes an integral with the measure $g(\lambda)$, the entanglement spectrum approaches the  delta-function, with the ``width'' of $g(\lambda)$ shrinking as $\Delta S$. We illustrate this behavior in Fig.~\ref{fig:renyi_to_spectrum1}, where we show qualitative 
index dependence of $S_A^n$ and corresponding entanglement spectrum $g(\lambda)$. In the companion holographic paper \cite{KdVETHholography}, we illustrate this behavior with an  explicit  toy model, exhibiting qualitatively similar properties. 

\section{Discussion}\label{sec:discussion}
In this paper we studied subsystem entropy for large subsystems of size $L_A$, which scale as a finite fraction of  the full system size,  $L_A=  x\, L,\, 0<x\leq 1$, in the thermodynamic limit $L\rightarrow \infty$. For 2d chaotic CFTs, presence of additional conserved quantum KdV charges implies that thermal equilibrium state should be locally described by the KdV generalized Gibbs ensemble (KdV GGE) \cite{Lashkari:2017hwq,Maloney:2018yrz,Dymarsky:2018lhf}. The Eigenstate Thermalization Hypothesis is also accordingly modified, with the CFT eigenstate being locally equivalent to the KdV GGE with the appropriate set of chemical potentials \cite{Dymarsky:2019etq}. We tested consistency of this picture by studying subsystem Renyi entropy when the full system is described by one of the following ensembles: canonical, microcanonical, primary eigenstate, and also the KdV GGE and generalized KdV-microcanonical ensemble. 

By focusing on the leading (extensive) contribution in the large volume limit $L\rightarrow \infty$, with the intensive properties of the state $\zeta$ kept fixed, we employed the semiclassical diagonal approximation of \cite{Dymarsky:2018}, discussed in section \ref{sec:semiclassical}, to evaluate Renyi entropy as a function of the subsystem volume $L_A=x\,L$, Renyi index $n$ and properties of the state $\zeta$,\footnote{Since we are only interested in the extensive part, we ignore, for example, the contributions of the order of $\sqrt{L_A}$, which appear when the subsystem is  as large as its compliment, $x=1/2$,  see \cite{PhysRevE.100.022131,Dong:2020iod}}.
\bea
\label{calSConclussions}
S_A^n=L\, {\cal S}(x,n,\zeta)+o(L_A).
\eea
We have found that our results, and in fact the intermediate field theory calculations based on the diagonal approximation, in the large $c$ limit match the holographic calculations of \cite{Dong:2018,Dong:2020iod,Dong:2023bfy,KdVETHholography}. Our results, naturally,  match all other known results for the subsystem entropy, previously discussed in \cite{Cardy:2014jwa} and many other papers. 
This further solidifies evidence behind the diagonal approximation of \cite{Dymarsky:2018}; also see \cite{Grover:2017,Jindal:2024zcg} for the effort to justify this approximation from the first principles. 

For convenience, we briefly reiterate the results for various ensembles, using the notation of \eqref{calSConclussions}, such that $x$-dependence of $\cal S$ defines the Page curve. 
Using (effective) temperature $\beta$ to characterize both canonical (Gibbs) and microcanonical ensembles, we found for the Gibbs ensemble
\be
\label{disc:Gibbs}
{\cal S}^{thermal}(x,n,\beta) = \left(\frac{c\pi}{6\beta}\right) x  \left(1+\frac{1}{n}\right), 
\ee
see \eqref{canonicalS} in section \ref{sec:Gibbs}, and 
\be
\label{disc:microcanonical}
{\cal S}^{mirco}(x,n,\beta) = \left(\frac{c\pi}{3\beta}\right)\frac{\sqrt{x+n^2(1-x)}-n}{1-n},
\ee
for the microcanonical ensemble, see \eqref{micro} in section \ref{sec:ordinary_micro}. As we emphasized in the main text, and was previously outlined for a general case in \cite{Dymarsky:2018,Grover:2018}, the subsystem entropy for the canonical and microcanonical ensembles would be the same 
only for von Neumann entropy $n=1$ or when the subsystem is very small $x\ll 1$. Although mean local properties  as measured in both ensembles are the same, main contribution to  entropy come from very different descendant states, characterized by \eqref{onshellq3} and \eqref{microq3} correspondingly.

Next we review the result for the KdV Generalized Gibbs Ensemble \eqref{kdvgge} with the chemical potentials (fugacities) $\mu_{2k-1}$. Its  subsystem entropy is given by 
\be
\label{disc:kdvgge}
{\cal S}^{GGE}(x,n,\mu_{2k-1}) = x \left[\frac{f(n \mu_{2k-1})-n f(\mu_{2k-1})}{1-n}\right],
\ee
where $f(\mu_{2k-1})$ is free energy density of the corresponding global state, see  \eqref{SGGE} in section \ref{sec:KdVGGE}. We note  that while \eqref{disc:kdvgge} exhibits the same linear subsystem size dependence as in the case of thermal ensemble, Renyi index dependence could be highly nontrivial. We  demonstrate that using several explicit examples in section \ref{sec:KdVGGE}. When $n=1$, i.e.~in the case of von Neumann entropy, at least in the leading order in $1/c$ \eqref{disc:kdvgge}  reduces to the conventional (micro)canonical result ${\cal S}=\frac{c\pi^2}{3\beta} x$.  

In addition to KdV GGE, we have considered generalized KdV-microcanonical ensemble, characterized by microcanonical window for several KdV charges $Q_{2k-1}$. Assuming thermodynamic limit, we 
define the ensemble as including equally-weighted CFT eigenstates with fixed charge densities $q_{2k-1}=Q_{2k-1}/L$, $k\leq m$, for some (small) $m$. Focusing on $m=2$ we found subsystem entropy explicitly at leading order in large central charge $c$: 
\bea
\label{disc:KdVmicro}
{\cal S}^{KdVm}(x,n,q_1,q_3)= \begin{cases}
\left(\frac{c\pi}{3\beta}\right)
\frac{\sqrt{x+n^2(1-x)}-n}{1-n},\qquad \, \,  \qquad\qquad\qquad \qquad \qquad \epsilon \geq  \epsilon_{crit},\\
\left(\frac{c\pi}{3\beta}\right)\frac{\sqrt{x\left[x-\sqrt{x(1-x)\epsilon}\right]}+n\sqrt{(1-x)\left[(1-x)+\sqrt{x(1-x)\epsilon}\right]}-n}{1-n},\;\;\epsilon \leq \epsilon_{crit}.
\end{cases}
\eea
Here charge densities $q_1,q_3$ are written in terms of the effective temperature $\beta=\sqrt{c\pi/12q_1}$ and $\epsilon=q_3/q_1^2-1$, and $\epsilon_{crit}(x,n)$ is defined in \eqref{epsiloncrit}. A novel feature in comparison with the conventional microcanonical ensemble is a sharp (at leading $1/c$ order) transition behavior. Considering subsystem Renyi entropy as a function of index $n$, with the subsystem size and the densities $q_1,q_3$ of the global state fixed, we find that for $n\leq n_{cirt}$, where 
\bea
\begin{aligned}
    n_{crit} &= \sqrt{q^{\bar{A}}_1/q^A_1},\quad 
    q^A_1 = q_1 \left(1-\sqrt{\frac{(1-x)\epsilon}{x}}\right), \quad
    q^{\bar{A}}_1 &= q_1 \left(1+\sqrt{\frac{x\epsilon}{1-x}}\right),\nonumber
\end{aligned}
\eea
Renyi entropy is the same as for the microcanonical ensemble \eqref{disc:microcanonical}. When $n$ increases and exceeds $n_{crit}$ the transition occurs, such that the main contribution to subsystem Renyi entropy is given by what can be interpreted as a tensor product of two fixed area states and corresponding  refined Renyi entropy becomes $n$-independent, see section \ref{sec:KdV-micro}. The same conclusions is reached holographically in \cite{KdVETHholography}. The same transition can be understood as a transition in $x$.  It  is illustrated in Fig.~\ref{fig:S2} in section \ref{sec:KdV-micro}.
When $c$ is finite, the sharp transition becomes a crossover. We illustrate that with numerical examples in the Appendix \ref{app:KdVmicro}.  

When the subsystem is small, $x\leq x_l$, subsystem entropy for the KdV-microcanonical ensemble is the same as for the microcanonical ensemble \eqref{disc:microcanonical}. The same  is also true for any $x$ when $n\rightarrow 1$. Thus, we find that when the subsystem is very small or for von Neumann entropy $n=1$ all four ``canonical'' ensembles (Gibbs, microcanonical, KdV GGE, and KdV microcanonical) would yield the same subsystem entropy at leading order in large $c$.

Presence of conserved KdV charges renders the conventional  Eigenstate Thermalization Hypothesis, that postulates equivalence between local properties of the (micro)canonical ensemble and a typical energy eigenstate, invalid. Instead the KdV-generalized version should take place \cite{Dymarsky:2019etq}, with the eigenstates labeled by charges $Q_{2k-1}$ being  locally equivalent to the KdV GGE or KdV microcanonical ensemble with matching parameters. The statement of generalized ETH, which we refer to as KdV ETH, can be extended to include non-local observables, such as subsystem entropy when $x<1/2$. In the latter case the eigenstate should be equated with the KdV microcanonical ensemble with matching charge densities. We test this extended version of ETH by considering primary  eigenstates, which are characterized by saturation of the inequalities \eqref{qineq} 
\be
q_{2k-1} =q_1^k.
\ee
In this case the diagonal approximation predicts an $n$-independent Renyi entropy 
\be
\label{disc:primary}
{\cal S}(x,n,\beta) = \frac{\sqrt{c(c-1)}\pi}{3\beta} x,
\ee
where we characterize the state by its  effective temperature $\beta=\sqrt{c\pi/12q_1}$.
The same result  was previously obtained at  the leading $c$ order in \cite{Wang:2018} by means of the monodromy method. As was emphasized in \cite{Dymarsky:2018lhf}, primary eigenstates sit at the boundary of the space of the allowed values of $Q_{2k-1}$. Therefore to  match such a state  by a KdV GGE, the latter has to have singular values of chemical potentials. We consider the appropriate limit of the KdV GGE with just two potentials $\mu_1,
\mu_3$ and match \eqref{disc:primary}  starting from \eqref{disc:kdvgge} in section \ref{sec:KdVGGE}. We note, that normally the GGE ensemble would only reproduce local properties of the eigenstates, hence matching the subsystem entropy of a primary state by a singular KdV GGE goes beyond the standard statement of the KdV ETH. To further test the KdV ETH we compare the primary state result \eqref{disc:primary} with that one of the KdV-microcanonical ensemble in the limit  $\epsilon\to 0$. Clearly in this limit \eqref{disc:KdVmicro} 
and  \eqref{disc:primary} agree within their rage of validity, i.e.~ at leading $1/c$ order. The comparison at finite $c$ is discussed in the Appendix  \ref{app:finite_c}. The general picture is as follows: when $\epsilon \ll 1$ index dependence of the subsystem Renyi entropy becomes suppressed with the difference
$\Delta S_A=S_A^{vN}-S_A^\infty$ becoming of order $\sqrt{\epsilon}\,  S_A^{vN}$. In terms of the entanglement spectrum, i.e.~the eigenvalues of $-\ln \rho_A$,  it means the latter becomes sharply peaked around $S_A^{vN}$ and approaches a delta-function in the limit $\epsilon\rightarrow 0$, matching flat entanglement spectrum of the primary state. We conclude by noting that agreement between the KdV microcanonical ensemble in the $\epsilon \rightarrow 0$ with the CFT primary states provides a consistency check for the KdV ETH and resolves the contradiction with the conventional ETH outlined in \cite{Wang:2018}. 

As we mentioned above, the KdV microcanonical ensemble matches conventional microcanonical ensemble in the $x\rightarrow 0$ limit, while also matching the primary state in the $\epsilon \rightarrow 0$ limit. We note that the subsystem Renyi entropy for these two cases \eqref{disc:microcanonical} and \eqref{disc:primary} are very different. This is possible because the  order of limits is important. At the leading order in $1/c$, when $\epsilon$ is small but fixed, as $x$ decreases and becomes smaller than $x_l(\epsilon,n)$ subsystem Renyi entropy is the same as for the microcanonical ensemble. On the other hand if $x$ is fixed while $\epsilon \rightarrow 0$, since for $\epsilon\ll 1$, $x_l\approx n^4/(n^2-1)^2\epsilon$, eventually $x$ will be large than  $x_l$. At that point  the subsystem Renyi entropy would match that one of primary state \eqref{disc:primary}, rather the than microcanonical result.

As we emphasized through the paper, the agreement between  primary states entropy with that one of the KdV microcanonical ensemble in an appropriate limit is a necessary condition for  KdV ETH. To fully test the hypothesis one should extend the analysis to highly excited descendant states, preferably with arbitrary values of $\epsilon=q_3/q_1^2-1$ (within the kinematic constraints). First steps in this direction were undertaken in \cite{Guo:2018,Brehm:2020zri}, which discussed entanglement entropy of various descendant states. Comparing these results with those for the KdV-charged ensemble is a task for the future. 
A related  generalization of this work would be to consider the KdV GGE with more than two chemical potential, or the KdV microcanonical ensemble with more than first two KdV charges fixed. A related holographic analysis based on two-zone solutions is partially carried out in \cite{KdVETHholography}. It would be interesting to leverage it to evaluate the subsystem entropy in the holographic limit, generalizing \eqref{disc:KdVmicro}.

Our results re-emphasize 2d CFTs as a potent playground to analytically study entanglement entropy of various ensembles, including those  that carry charges with respect to enhanced symmetry. So far exact results for eigenstates or charged ensembles in interacting integrable models with additional symmetries were scarce \cite{Zhang:2021bmy}.
Inspired by the seminal work  of Calabrese  and Cardy \cite{Calabrese:2004eu,Calabrese:2005in}, which influenced many analytic and numerical studies of entanglement, we hope our results for the KdV-charged ensembles will motivate further studies of entanglement in integrable spin chains. 
Our results is also an illustration of the power of the diagonal approximation, which can be used  beyond the CFT settings. We have seen that the diagonal approximation mirrors, on the field theory side, the  holographic calculations of \cite{Dong:2018,Dong:2020iod,Dong:2023bfy,KdVETHholography}. 
It would be interesting to apply this technique beyond conformal field theories, to study entanglement e.g.~in various lattice models. 

Another lesson highlighted by our results is that Renyi entropy is a sensitive probe of the underlying quantum state. The von Neumann entropy is known to be blind to many aspects of the global state, and we saw that subsystem von Neumann entropy for  many ensembles is the same. This  echos numerical studies of the eigenstate subsystem entropy in integrable spin chains at criticality \cite{PhysRevLett.127.040603,PhysRevA.104.022414,Miao2022eigenstate}. In all of these cases weak ETH is sufficient to ensure that eigenstate von Neumann entropy follows the Calabrese-Cardy analytic result with a spectacular precision. We emphasize that repeating similar studies in e.g.~XXZ chain for Renyi entropy with a large index $n$ would require going beyond weak ETH, and in fact would provide a sensitive test of generalized ETH  \cite{Vidmar_2016} (the analog of the KdV ETH for the spin chain). We outline this as an important task for the future.

\section*{Acknowledgments}
We thank Xi Dong and Lev Vidmar for useful discussions. L.C and H.W are supported by National Science Foundation of China (NSFC) grant no. 12175238. JT is supported by  the National Youth Fund No.12105289. AD is supported by the NSF under grant PHY-2310426. This work was performed in part at Aspen Center for Physics, which is supported by National Science Foundation grant PHY-2210452.

\appendix

\section{Free energy of KdV GGE}
\label{app:f}
Following \cite{Dymarsky:2018iwx,Dymarsky:2018lhf,Dymarsky:2019etq} in this section we review large $c$ limit of  the  free energy expression for the KdV GGE  with only two chemical potentials $\mu_1,\mu_3$ turned on 
\bea
e^{F}={\rm Tr}\,e^{-\mu_1 Q_1-\mu_3 Q_3}.
\eea
Here we absorb $-\beta$ into the definition of $F$. 
For convergence it is necessary and sufficient to require $\mu_3>0$ while $\mu_1$ could be arbitrary. We introduce, c.f~\eqref{tau}, 
\bea
\label{app:tau}
\tau={\mu_1\over \mu_3^{1/3}}\left(12\over c\pi\right)^{1/3}.
\eea
Assuming thermodynamic limit, we are interested in free energy density $f=F/L$, c.f~\eqref{ftau},
\bea
f=\left({c\pi \over 12 }\right)^{2/3} \mu_3^{-1/3}(f_0(\tau)+f_1(\tau)/c+O(1/c^2)).
\eea
Holographically  $f_0$ is the contribution of the BTZ black hole (there are no other classical saddles in this case  \cite{Dymarsky:2020}), and $f_1$ is a contribution of small fluctuations around the BTZ background. Explicitly, 
\bea
f_0(\tau)=2\sqrt{\sigma_0}-{\tau}\sigma_0-\sigma_0^2,
\eea
where function $\sigma_0(\tau)$ satisfies $\sqrt{\sigma_0}={\tau}\sigma_0+2\sigma_0^2$ and is given explicitly by 
\bea
\label{sigma0}
\sigma_0(\tau):=\frac{\left(\sqrt[3]{\tau ^3+3 \left(\sqrt{6 \tau ^3+81}+9\right)}-\tau \right)^2}{6 \sqrt[3]{\tau ^3+3 \left(\sqrt{6 \tau ^3+81}+9\right)}}.
\eea
Function $f_1$ is more elaborate
\bea
f_1(\tau)&=&-\sqrt{\sigma_0}\left(1+{12\over \pi}\int_0^\infty d\kappa \ln \left[1-e^{-2\pi \kappa \gamma}\right]\right),\\
\gamma&=&1+4(1+\kappa^2)\sigma_0^{3/2}=\\
&&1+\frac{\sqrt{\frac{2}{3}} \left(\kappa ^2+1\right) \left(\sqrt[3]{\tau ^3+3 \left(\sqrt{6 \tau ^3+81}+9\right)}-\tau \right)^3}{3 \sqrt{\tau ^3+3 \left(\sqrt{6 \tau ^3+81}+9\right)}}.
\eea
From here we readily find \cite{Dymarsky:2019etq},
\bea
q_1&=&-{\partial f\over \partial \mu_1}=\left({c\pi\over 12}\right)^{1/3} \mu_3^{-2/3}\left(\sigma_0-{1\over c}{\partial f_1\over  \partial \tau}+O(1/c^2)\right),\\
q_3&=&-{\partial f\over \partial \mu_3}=\left({c\pi\over 12}\right)^{2/3} \mu_3^{-4/3}\left(\sigma_0^2+{1\over 3 c} \left(f_1+\tau {\partial f_1\over \partial \tau}\right)+O(1/c^2)\right),
\eea
and, 
\bea
c\, \epsilon&=&c\left({q_3\over q_1^2}-1\right)=y(\tau)+O(1/c),\\
y(\tau)&=&96\int_0^\infty {d\kappa \kappa(1+\kappa^2)\over e^{2\pi \kappa \gamma}-1}.
\label{yfunction}
\eea
\begin{figure}
    \centering
    \includegraphics[width=0.6\textwidth]{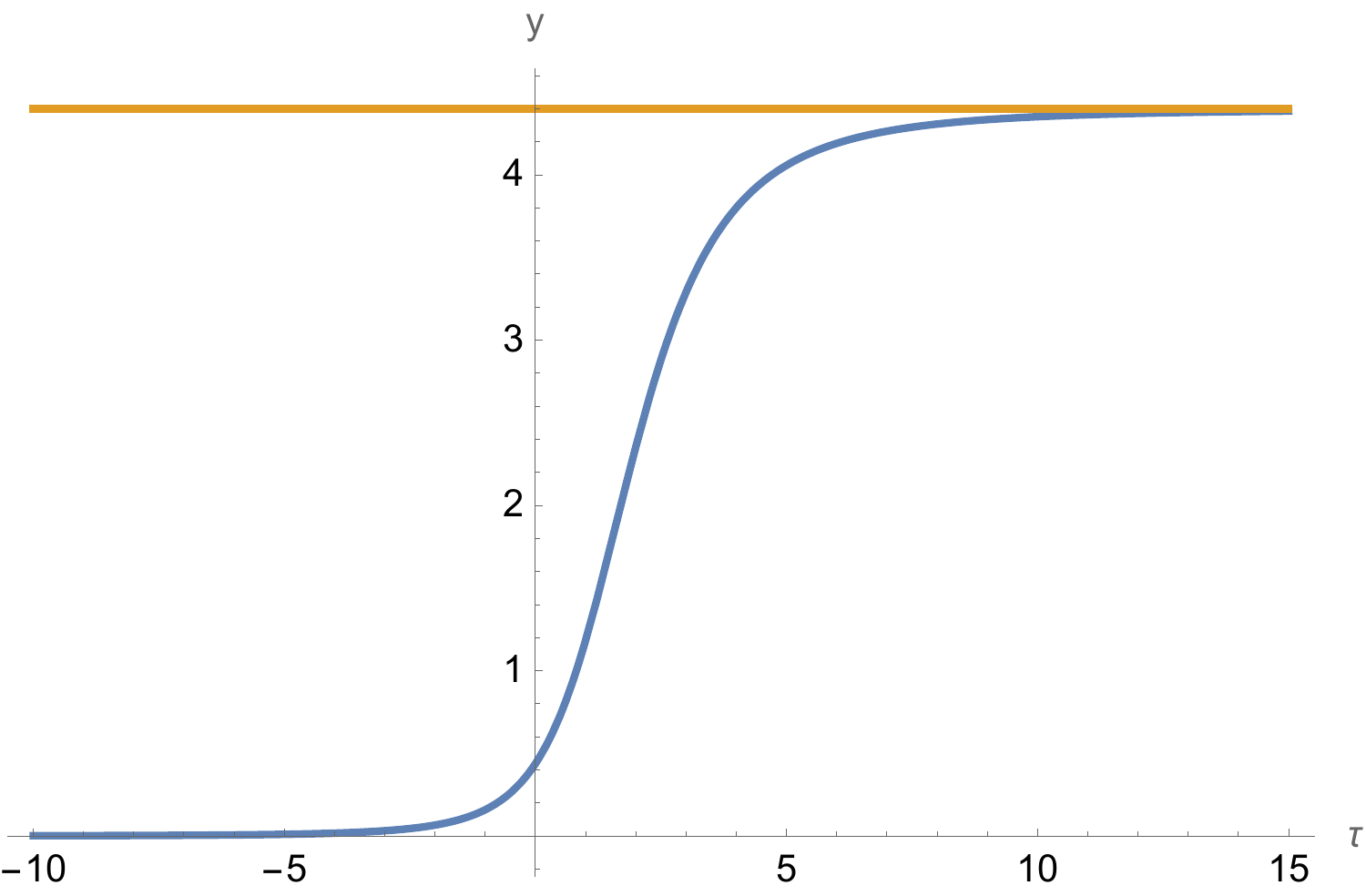}
    \caption{Plot of function $y(\tau)$ (blue) defined in \eqref{yfunction}. Gibbs ensemble value $y=22/5$ is shown in orange.}
    \label{fig:y}
\end{figure}
The plot of function $y(\tau)$ is shown in Fig.~\ref{fig:y}. It interpolates between the Gibbs ensemble value $y=22/5$ for $\tau\rightarrow\infty$ and primary state value $y=0$ for $\tau\rightarrow -\infty$. Energy density can be express in terms of $y(\tau)$ as follows,
\bea
q_1&=&\left({c\pi\over 12}\right)^{1/3} {\sigma_0\over \mu_3^{2/3}} \left(1+{f_1-3\sigma_0^2 y\over c \sigma_0(\tau+6\sigma_0)}+O(1/c^2)\right),\\
\eea
For $\tau=0$ we find
\bea
\sigma_0(0)=2^{-2/3},\quad f_0(0)= 2^{-4/3}3,\quad f_1\approx -0.53092,\quad y(0)\approx 0.43724,
\eea
and  
\bea
q_1=\left({c\pi\over 12}\right)^{1/3}{1\over 2^{2/3}\mu_3^{2/3}}\left(1-{0.441592\over c}+O(1/c^2)\right)={c\pi\over 12\beta^2},
\eea
where $\beta(\mu_3)$ is the effective temperature.
Combining this together we arrive at \eqref{st0}.

In the limit $\tau\rightarrow -\infty$ we find 
\bea
\sigma_0={|\tau|\over 2}+{1\over \sqrt{2|\tau|}}+\dots,
\eea
and 
\bea
q_1={-\mu_1\over 2\mu_3}(1+o(1/c^2)),
\eea
with the free energy density given by \eqref{primaryf}.

\section{Reverse engineering $s(q_1,q_3)$} 
\label{app:s}
In what follows we will try to reverse-engineer the KdV microcanonical entropy $S(Q_1,Q_3)$ in the thermodynamic limit when both charges scale linearly with the system size $L$. We thus can assume 
\bea
S(Q_1,Q_3)=L s(q_1,q_3),
\eea
where $q_i$ are charge densities. It is convenient to introduce $x\geq 0$,
\bea
\label{defx}
q_3=q_1^2(1+\epsilon),
\eea
such that 
\bea
s(q_1,q_3)= \sqrt{c\pi q_1\over 3} s(\epsilon,c).
\eea
Dependence on $q_1$ is fixed by the requirement that $S(Q_1,Q_3)$ is extensive. 
We have also emphasized explicit $c$ dependence of $s$. 

There is a number of consistency conditions that function $s(x,c)$ must satisfy. First, upon integration over $dq_3$ with fixed $q_1$ we expect 
\bea
\int dq_3\, e^{L {\cal S}(q_1,q_3)}=q_1^2\int d\epsilon\, e^{L \pi \sqrt{2c q_1\over 3} s(x,c)}=e^{L \pi \sqrt{2c q_1\over 3}}.
\eea
Since in the large $L$ limit an integral can be evaluated using saddle point approximation, we conclude that at the maximum $\epsilon^*$ where $ds/d\epsilon=0$ the value of $s(\epsilon^*,c)=1$ for any value of $c$.
We also know that the mean value of $q_3$ in the (micro)canonical ensemble is $q_3=q_1^2\left(1+{22\over 5c}\right)$. We thus conclude that for any value of $c$, $s(x,c)$ reaches its maximum at 
\bea
\epsilon^*={22\over 5c},\qquad \left.{ds\over d\epsilon}\right|_{\epsilon^*}=0,\qquad s(\epsilon^*,c)=1.
\eea

A more stringent set of constraints comes from matching KdV GGE free energy
\bea
e^{F}=L^2\int dq_1\, dq_3\, e^{L (\sqrt{c\pi  q_1\over 3} s(x,c)-\beta q_1-\mu_3 q_3)}.
\eea
The extensive part of free energy $F=L\, f+o(L)$ is given by the saddle point
\bea
\label{f13}
f(q_1,\epsilon)=\pi \sqrt{2c q_1\over 3} s(\epsilon,c)-\beta\, q_1-\mu_3\, q_1^2(1+\epsilon),
\eea
where $q_1,\epsilon$ satisfy $\partial_{q_1}f(q_1,\epsilon)=\partial_\epsilon f(q_1,\epsilon)=0$.
We will first solve these equations perturbatively around the ``canonical ensemble'' point $\mu_3=0$. To that end we introduce $t_3={c\pi^2 \mu_3\over 6\beta^3}$ and will expand everything in powers of $t_3$. 
First we introduce an auxiliary algebraic equation 
\bea
{d (2\sqrt{\sigma}-\sigma-t_3\, \sigma^2)\over d\sigma}=0,
\eea
with the solution 
\bea
\sigma(t_3)=1-4t_3+28t_3^2+\dots,
\eea
Then we solve saddle point equations $\partial_{q_1}f(q_1,\epsilon)=0, \partial_{\epsilon}f(q_1,\epsilon)=0$, 
\bea
\label{sp1}
q_1&=&{c\pi\over 12\beta^2}s^2(\epsilon,c) \sigma(\tilde{t}_3),\quad \tilde{t}_3\equiv t_3 s^2(\epsilon,c)(1+\epsilon),\\ 
\label{sp2}
{ds\over d\epsilon}&=&{s(\epsilon^*,c)\over 2(1+\epsilon)} \tilde{t}_3\sigma^{3/2}(\tilde{t}_3).
\eea
Free energy can be written as 
\bea
F={c\pi L\over 12\beta}f,\qquad 
f=s^2(\epsilon,c) (2\sqrt{\sigma}-\sigma-\tilde{t}_3 \sigma^2),
\eea
where the argument of $\sigma$ is $\tilde{t}_3$. 
We now compare this result with known large $c$ expansion of free energy,
\bea
\label{freeeexp}
f=f_0+f_1/c+f_2/c^2+\dots,
\eea
where $f_0=(2\sqrt{\sigma}-\sigma-t_3 \sigma^2)$ and here $\sigma=\sigma(t_3)$. From here we conclude that 
\bea
s(x,c)=1-a_2\, c\,\left(x-{22\over 5c}\right)^2+ a_3\, c^2\left(x-{22\over 5c}\right)^3+a_4\, c^3\left(x-{22\over 5c}\right)^4+\dots,
\eea
where all $a_i$ are functions of $c$, starting from $c^0$ order in $1/c$ expansion. At the leading order  in large  $c$ limit, 
the difference $\epsilon-22/(5c)$ will be of order $1/c$, and $s(\epsilon,c)=1+O(1/c)$. In fact, since $f_2$ in \eqref{freeeexp} starts with $t_3^2$, we find exact value of
\bea
\label{alpha}
a_2=\frac{175 c}{8 (4320 c-13552)}={35\over 6912}+O(1/c).
\eea
In the same way we can also fix the leading order of 
\bea
a_3=\frac{8575}{8957952}+O(1/c).
\eea
This finishes the derivation of \eqref{spert}.

Next we proceed to  fix first subleading order $s_1$ of $s=1-s_1/c+\dots$. To that end we once again consider \eqref{f13} with positive  $\mu_3$ and arbitrary $\beta$. As was discussed in Appendix \ref{app:f}, free energy is a function of $\tau$ defined in \eqref{app:tau},
which can take arbitrary real values. The ``on-shell'' value of $y= c\, \epsilon$ at leading $1/c$ order is given by \eqref{yfunction}, with large $\tau$ asymptotics  are as follows 
\bea
y&=&{22\over 5}-\frac{1728}{35\tau^3}+\dots,\qquad \tau\rightarrow+\infty,\\
y&=&{2\over (-\tau)^3}+\dots,\qquad \qquad\tau\rightarrow -\infty.
\eea

First, in full generality, i.e. without expanding in $1/c$, the extensive part of free energy \eqref{f13}  can be written as 
\bea
f=\left({c\pi s^2\over 12 }\right)^{2/3} {1\over (\mu_3(1+\epsilon))^{1/3}}(2\sqrt{\sigma_0(\tilde{\tau})}-\tilde{\tau}\sigma_0(\tilde{\tau})-\sigma_0^2(\tilde{\tau})),
\eea
where
\bea
q_1=\left({c\pi s^2\over 12\mu_3^2 (1+\epsilon)^2 }\right)^{1/3}\sigma_0(\tilde{\tau}),\qquad \tilde{\tau}={\tau \over (1+\epsilon)^{1/3}s^{2/3}}.
\eea
Function $\sigma_0$ above is the same as in \eqref{sigma0}, while 
$s$ as a function of $\epsilon$ satisfies 
\bea
\label{sdiffeq0}
2{1+\epsilon\over s}{ds\over dx}={f_0(\tilde{\tau})+\tilde{\tau} f_0'(\tilde{\tau})\over 2f_0(\tilde{\tau})-\tilde{\tau} f_0'(\tilde{\tau})},\qquad f_0(\tau)=2\sqrt{\sigma_0(\tau)}-\tau \sigma_0(\tau)-\sigma_0^2(\tau).
\eea
This follows from saddle point equations $\partial_{q_1}f=\partial_\epsilon f=0$. 
Equation \eqref{sdiffeq0} can be further simplified as 
\bea
\label{sdiffeq}
{d\ln s\over d\ln(1+\epsilon)}={\sigma^{3/2}_0(\tilde{\tau})\over 2}.
\eea
Given that $\tilde{\tau}$ is a function of $\epsilon$ and $s(\epsilon)$, equation \eqref{sdiffeq} becomes a very complicated differential equation to solve, even if we know $\epsilon(\tau)$ from the (hypothetical) exact result for the KdV GGE free energy.

Things simplify substantially at the leading $1/c$ order. We can assume that 
\bea
s(\epsilon,c)=1-{s_1(y)\over c}+O(1/c^2)
\eea
where $\epsilon=y/c$ and $s_1(y)$ is some $c$-independent function. We then have
\bea
\label{syequation}
-{ds_1\over dy}={\sigma^{3/2}_0(\tau)\over 2},
\eea
and $y(\tau)$ is given by \eqref{yfunction}. We also take into account that when $\beta>0,\mu=0$, which correspond to $\tau\rightarrow \infty$, we have $y=22/5$ and $s_1(22/5)=0$. Using large $\tau$ asymptotic of $\sigma_0(\tau)\approx 1/\tau^2$, we find 
\bea
s'_1(y)\approx {1\over 2\tau^3}\approx-{35\over 3456}\left(y-{22\over 5}\right)\quad \Rightarrow\quad  s_1=-{\frac{35}{6912}}\left(y-{22\over 5}\right)^2+\dots 
\label{qc}
\eea
This matches the coefficient in \eqref{alpha}, as expected.

An analogous analysis for $\tau\rightarrow -\infty$ suggest 
\bea
\sigma_0(\tau)\approx {|\tau|\over 2}\approx {1\over 2}\left({2\over y}\right)^{1/3},
\eea
from where we find for small $y\rightarrow 0$
\bea
\label{smally}
-s_1={\sqrt{y}\over 2}-{\rm s}_1,
\eea
where the constant $c$ has to be  chosen such that $s_1(22/5)=0$. Numerically it is very close to ${\rm s}_1=1/2$. This matches analytic value ${\rm s}_1=1/2$, which can be derived from \eqref{eq:Cardy_corrected} by noting that $\epsilon\rightarrow 0$ is the limit when $n=0$ and $\Delta= L\, Q_1$. Hence in this limit 
\bea
s(0,c)=\sqrt{c-1\over c}=1-{1\over 2c}+\dots.
\eea

By numerically integrating \eqref{syequation} we find the profile  of $s_1(y)$, shown in Fig.~\ref{fig:s1}.
Finally, we compare this result with the naive approximation  \eqref{naive}.
We notice that 
small $y$ asymptotic \eqref{smally} is the same, while the leading coefficient of the Taylor series expansion near $y=22/5$ \eqref{qc} is different: $-\frac{25}{4608}$ vs $-\frac{35}{6912}$.  Nevertheless the exact and approximate results are numerically close, with the corresponding  curves visually indistinguishable in Fig.~\ref{fig:s1}.  

\section{Large $c$ limit of $s(q_1,q_3)$}
\label{app:shol}
In the ``holographic'' limit of large $c$ we assume scaling $q_1\propto c$, $q_3\propto c^2$ and  $c$-independent $\epsilon$. In this limit quantum KdV hierarchy becomes classical; instead of quantum states one considers the so-called finite zone solutions, see \cite{Novikov:1974,Dymarsky:2020,Dymarsky:2022dhi}. Classical KdV charges are then given by 
\bea
Q_1^{cl}=\left({12 L\over c\pi}\right) Q_1&=&h+\sum_k k\, {\mathcal I}_k,\\ \label{expansionQ}
Q_3^{cl}=\left({12 L\over c\pi}\right)^2 Q_3&=&h^2 +\sum_k k(4k^2+6h)\, {\mathcal I}_k+O({\mathcal I}^2),\\
\dots
\eea
where $h$ is the invariant of the co-adjoint orbit of Virasoro algebra and ${\mathcal I}_k$ are action variables. In the semiclassical limit $h$ is related to primary dimension $\Delta$ and action variables are related to quantum numbers $m_k$ in \eqref{Virstate} \cite{Dymarsky:2022dhi}. 

Holographically, entropy is given by the black hole horizon area, which in case of finite zone solutions is given by \cite{Dymarsky:2020}, 
\bea
S={c\pi\over 6}\sqrt{h}. \label{shol}
\eea
We are focusing on one zone-solutions, when only one ${\mathcal I}_k$ for a particular $k$ is non-zero.
Working in the limit of small ${\mathcal I}_k\ll 1$, such that \eqref{expansionQ} is reliable, which is the limit of $\epsilon=q_3/q_1^2-1$ of order one in terms of $1/c$ expansion but numerically small, we find 
\bea
\epsilon={4k(k^2+h)\over h^2}{\mathcal I}_k+O({\mathcal I}_k^2),
\eea
and 
\bea
h=Q_1^{cl}\left(1-{Q^{cl}_1\, \epsilon\over 4(k^2+Q^{cl}_1)}+O(\epsilon^2)\right).
\eea
For any fixed $k$, in the thermodynamic limit we find $s(\epsilon)=1-\epsilon/8+O(\epsilon^2)$. This result applies to a particular family of one-zone solutions with the given $k$. To find microcanonical entropy with fixed $Q_1,Q_3$ we should maximize \eqref{shol} with respect to $k$ first and  take $Q_1\rightarrow \infty$ afterwords. This is the limit when   $k$ goes to infinity faster than $Q_1$ and hence $s(\epsilon)=1$ for any $\epsilon$. In other words, holographic approximation yields $\epsilon$-independent entropy for $\epsilon$ of order $c^0$, which is consistent with  \eqref{eq:holo_entropy}.

A non-perturbative derivation of $\epsilon$-independence of entropy at leading $1/c$ order, not limited to the regime of small ${\mathcal I}_k\ll 1$, can be found in \cite{KdVETHholography}.

\section{Subsystem Renyi entropy of KdV microcanonical ensemble}
\subsection{General properties at finite $c$}
\label{app:finite_c}
We can begin by exploring the behavior of the subsystem Renyi entropy based on the saddle point equations 
\be\label{eq:micro_saddle}
\partial_{1} s(q^A_1,q^A_3) = n\, \partial_1 s(q^{\bar{A}}_1,q^{\bar{A}}_3),\qquad \partial_3 s(q^A_1,q^A_3) = n\,\partial_3 s(q^{\bar{A}}_1,q^{\bar{A}}_3)
\ee
derived from (\ref{effa}) amended by the the KdV charges conservation. Here $\partial_1$ and $\partial_3$ stand for the derivative of first and second argument of $s(q_1,q_3)$ correspondingly. 
We are particularly interested in the asymptotic behaviors of $S^n_A(q_1,q_3)$ in the limits $n\to 1$ and $n\to \infty$. The first limit is that one of the von Neumann entropy; in full generality entropy in this case is given by the KdV microcanonical thermal entropy 
\be
\lim_{n\to 1} S^n_A(q_1,q_3) = S^{vN}_A(q_1,q_3) = L_A\, s(q_1,q_3).
\ee
To investigate the other limit, we use the proposed form of the entropy function 
\be 
s(q_1,q_3) = \sqrt{\frac{c\pi q_1}{3}} s(\epsilon,c)
\ee
in conjunction with the saddle-point equations (\ref{eq:micro_saddle}). After some algebraic massaging, we arrive at the following equations
\bea\label{eq:saddle_ansatz} 
\sqrt{\frac{q^A_1}{q^{\bar{A}}_1}} \left[\frac{s(\epsilon^{\bar{A}},c)-4s'(\epsilon^{\bar{A}},c)(1+\epsilon^{\bar{A}})}{s(\epsilon^A,c)-4s'(\epsilon^A,c)(1+\epsilon^A)}\right] =\frac{1}{n},\;\;\; \left(\frac{q^A_1}{q^{\bar{A}}_1}\right)^{3/2} \left[\frac{s'(\epsilon^{\bar{A}},c)}{s'(\epsilon^A,c)}\right] = \frac{1}{n}
\eea
where we have denoted
\be 
s'(\epsilon,c) \equiv \partial_\epsilon s(\epsilon,c),\qquad \epsilon^{A,\bar{A}} = q^{A,\bar{A}}_3/\left(q^{A,\bar{A}}_1\right)^2-1.
\ee
Assuming that $q^{A,\bar{A}}_1$ are non-vanishing, the only way to solve the saddle-point equation consistently in the asymptotic limit $n\to \infty$ is to require that
\be
\lim_{n\to \infty}s'(\epsilon^A,c) \to \infty.
\ee
For a function $s(\epsilon,c)$ that is smooth for $\epsilon\geq 0$ as was proposed, its derivative $s'(\epsilon,c)$ can diverge only at the boundary, i.e. $\epsilon\to 0$ or $\epsilon \to \infty$. The total charge conservation dictates that $\epsilon^A$ must be bounded from above. We are therefore left with the only possibility that
\be
\lim_{n\to \infty}\epsilon^A \to 0.
\ee
The derivative of $s'(\epsilon,c)$ is indeed expected to diverge  at $\epsilon=0$, as follows from the behavior  $s\propto \sqrt{\epsilon}$ for small $\epsilon$, see section \ref{SQ1Q3}. We therefore conclude that the asymptotic saddle-point solution of \eqref{eq:micro_saddle} in the limit $n\to \infty$ saturates
\be
q^A_3 \to (q^A_1)^2.
\ee
Fixing this for the subsystem $A$, we can take the ratio of (\ref{eq:saddle_ansatz}) in conjunction with the charge conservation. Doing this produces the following equation that determines the remaining variable $\epsilon^{\bar{A}}$
\be
\frac{s'(\epsilon^{\bar{A}},c)}{s(\epsilon^{\bar{A}},c)} = \frac{x\left(1-x+\sqrt{x(1-x)\left(\epsilon-(1-x)\epsilon^{\bar{A}}+x\, \epsilon\, \epsilon^{\bar{A}}\right)}\right)}{4(1+x\,\epsilon^{\bar{A}})\sqrt{x(1-x)\left(\epsilon-(1-x)\epsilon^{\bar{A}}+x\, \epsilon\, \epsilon^{\bar{A}}\right)}},
\ee
where we remind that $\epsilon = q_3/q_1^2 -1$. The exact solution for $\epsilon^{\bar{A}}$ depends on the explicit form of $s(\epsilon,c)$. In the limit of small $\epsilon\ll 1$, it follows  from the divergence  of $s'(\epsilon^{\bar{A}},c)$ for $\epsilon^{\bar{A}} \rightarrow 0$ there is a solution for  $\epsilon_A$, which satisfies 
\be
\label{smalleplimit}
\epsilon^{\bar{A}} \propto \epsilon,\quad q^{\bar{A}}_1 -q_1 \propto q_1 \sqrt{\epsilon}.  
\ee
Assuming this is the dominant saddle-point solution, the asymptotic Renyi entropy can then be expressed in terms of the corresponding asymptotic solutions $q^{\bar{A}}_1, q^{\bar{A}}_3$ as follows 
\bea 
\lim_{n\to \infty} S^n_A(q_1,q_3) &=& L(s(q_1,q_3)-(1-x) s(q^{\bar{A}}_1,q^{\bar{A}}_3))\nonumber\\
&=& \lim_{n\to 1} S^n_A(q_1,q_3) +\Delta S_A(q_1,q_3),\nonumber\\
\Delta S_A(q_1,q_3)&=&L (1-x) \left(s(q_1,q_3)-s(q^{\bar{A}}_1,q^{\bar{A}}_3)\right).
\eea
It therefore follows from \eqref{smalleplimit} that the relative gap between the asymptotic values is also suppressed in the limit of $\epsilon \ll 1$, 
\be
\Delta S_A(q_1,q_3)/S^{vN}_A(q_1,q_3)\propto \sqrt{\epsilon}. 
\ee
This is a universal property of the subsystem Renyi entropy of  the KdV micro-canonical ensemble in the limit $\epsilon \ll 1$. It is valid for any $c$ and relies only on the qualitative behavior  of $s(\epsilon,c)$. 

\subsection{Computation of Renyi entropy using approximate  $s(q_1,q_3)$}
\label{app:KdVmicro}
\subsubsection{Renyi entropy at finite $c$}
We illustrate general properties of  the  subsystem Renyi entropy for the KdV microcanonical ensemble by performing explicit computations using the approximate expression (\ref{eq:KdV_entropy}) for $s(q_1,q_3)$. We first compute the results at finite $c$ of order one, and then consider $c\gg 1$. 

We are in particular interested in computing the Renyi entropy dependence on the index $n$. In full generality, the value of $S_A^n$ near $n=1$ is the same  as for the micro-canonical ensemble. The behavior for large  $n$ is less trivial. We can probe it by solving the saddle-point equations perturbatively in $1/n$. It can be shown that the subsystem charge densities $q^A_1, q^A_3 = (q^A_1)^2(1+\epsilon_A)$ as well as $q^{\bar{A}}_1, q^{\bar{A}}_3=(q^{\bar{A}}_1)^2\left(1+\epsilon_{\bar{A}}\right)$ admit the following expansion 
\bea\label{eq:large_n_asymptotic}
q^A_1 &=& q_1\left(\alpha_0 + \alpha_1 n^{-1}+... \right),\;\;\epsilon_A = \gamma_2\, n^{-2}+...\nonumber\\
q^{\bar{A}}_1 &=& q_1\left(\beta_0 + \beta_1 n^{-1}+... \right),\;\;\epsilon_{\bar{A}} = \kappa_0+\kappa_1 n^{-1}+...
\eea
We observe that $\epsilon_A$ does decay as $n^{-2}$ while $\epsilon_{\bar{A}}$ approaches a finite value $\kappa_0$ in the large $n$ limit. This is consistent with our general consideration in Appendix \ref{app:finite_c} above.

The expansion coefficients $\alpha_i,\beta_i,\gamma_i,\kappa_i$ are complicated functions of $c$ and $\epsilon$ that are not particularly important. In the limit of $\epsilon \ll 1$, one can show that 
\be
\lim_{n\to \infty} \epsilon_{\bar{A}}=\kappa_0 = \frac{\epsilon(1-x)^{-1}}{1+(c-1)x}+...
\ee
which leads to $\Delta S_A=S_A^{\infty}-S_A^1\propto \sqrt{\epsilon}$, as we discussed in Appendix \ref{app:finite_c}. In Figure (\ref{fig:Renyi_num}) we present a series of plots from a numerical computation that demonstrates the main features discussed in (\ref{app:finite_c}) explicitly. 

\begin{figure}[ht]
    \centering
    \includegraphics[width=0.32\linewidth]{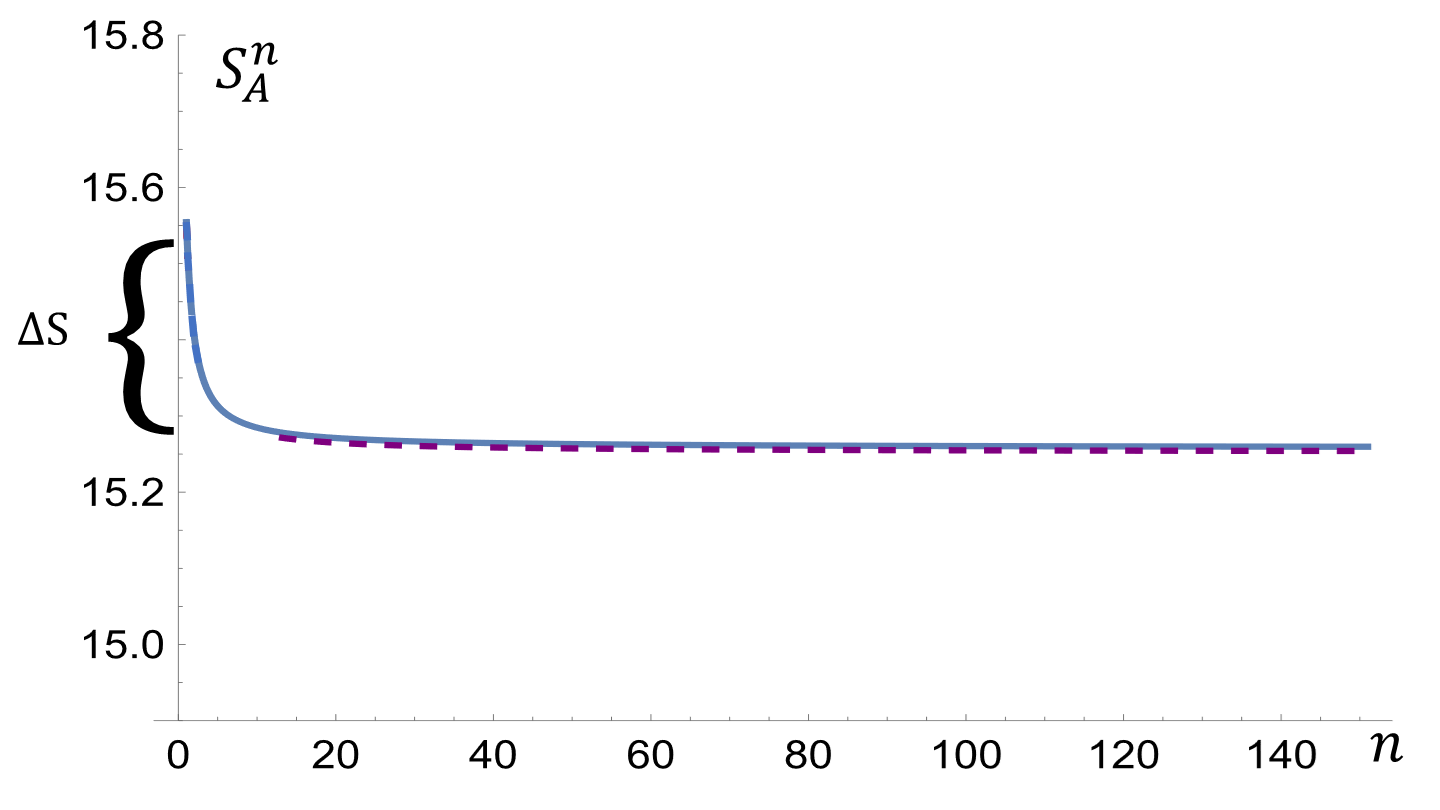}
    \includegraphics[width=0.32\linewidth]{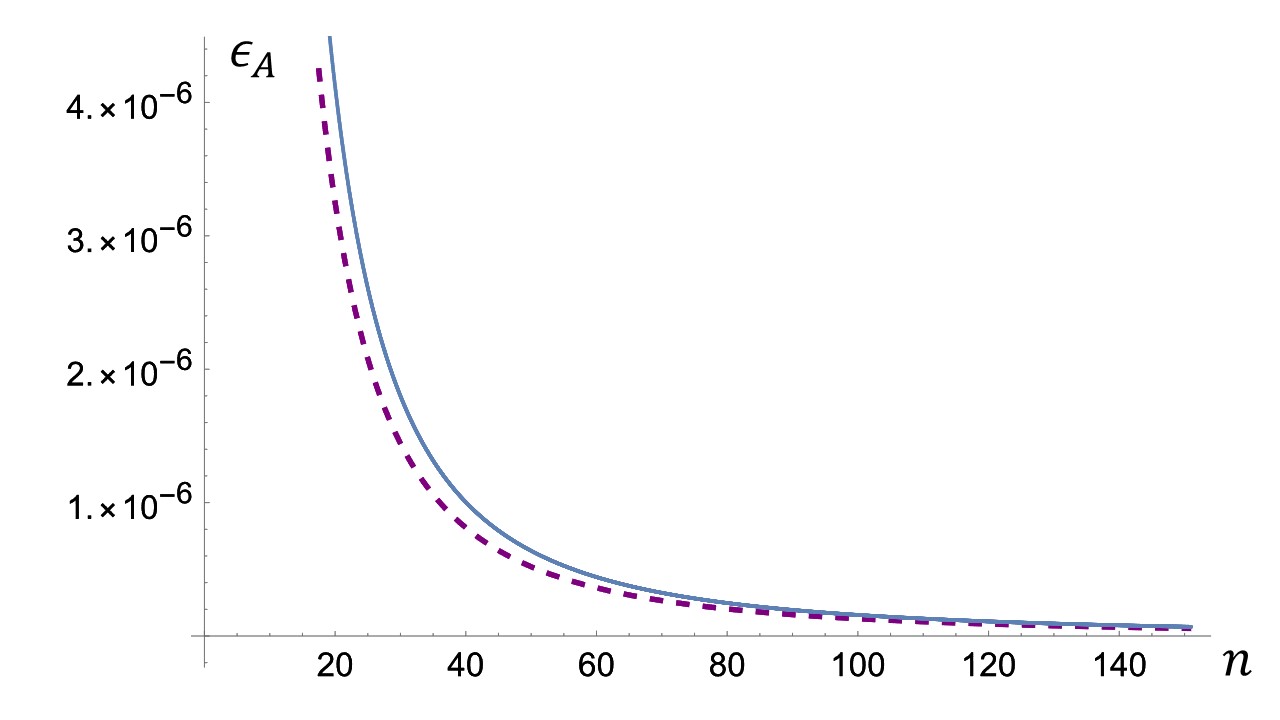}
    \includegraphics[width=0.32\linewidth]{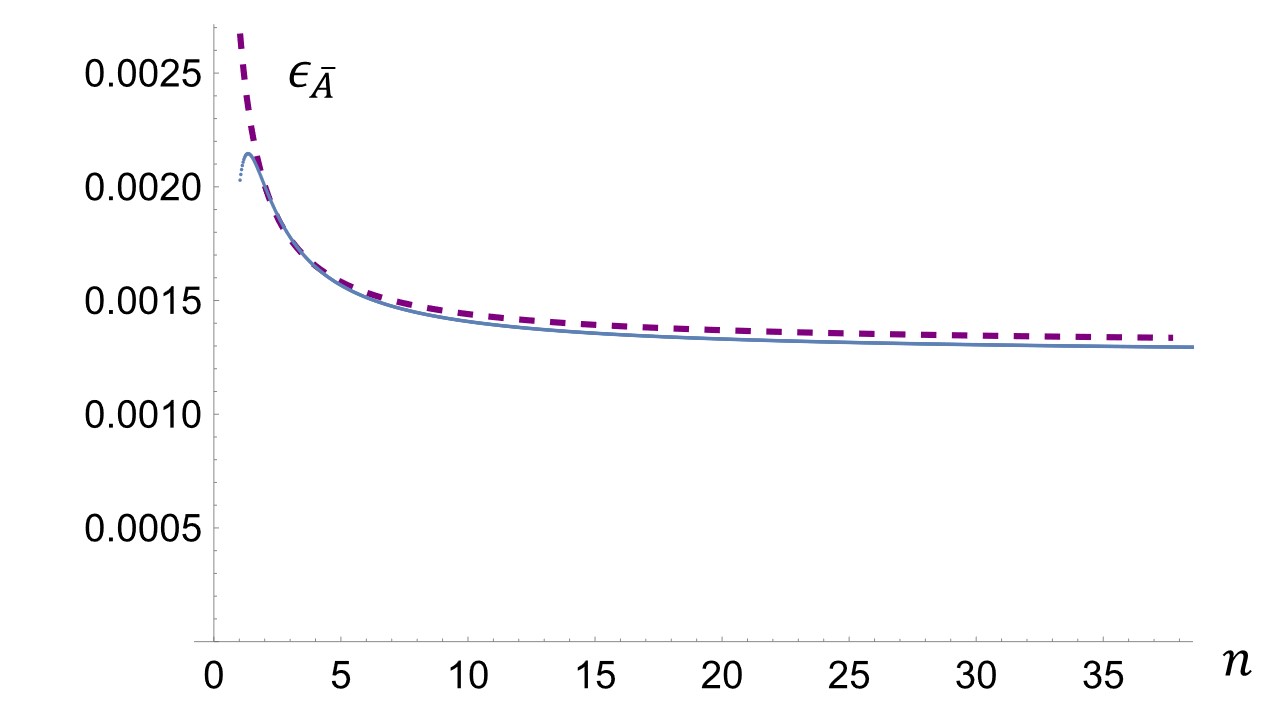}
    \caption{Numerical results, obtained using the approximate expression \eqref{eq:KdV_entropy}, shown in solid blue, against large $n$ asymptotic expression  (\ref{eq:large_n_asymptotic}) shown in dashed purple, for: subsystem Renyi entropy $S_A^n$ (left panel),   $\epsilon_{\bar{A}}$ (central panel), $\epsilon_A$ (right panel). The parameters are chosen as follows: $x=0.3, c=5, q_1 = 20 c, \epsilon = 10^{-2}/c$. }
    \label{fig:Renyi_num}
\end{figure}

\subsubsection{Renyi entropy in the holographic limit $c\to \infty$}
\begin{figure}[ht]
    \centering
    \includegraphics[width=0.48\linewidth]{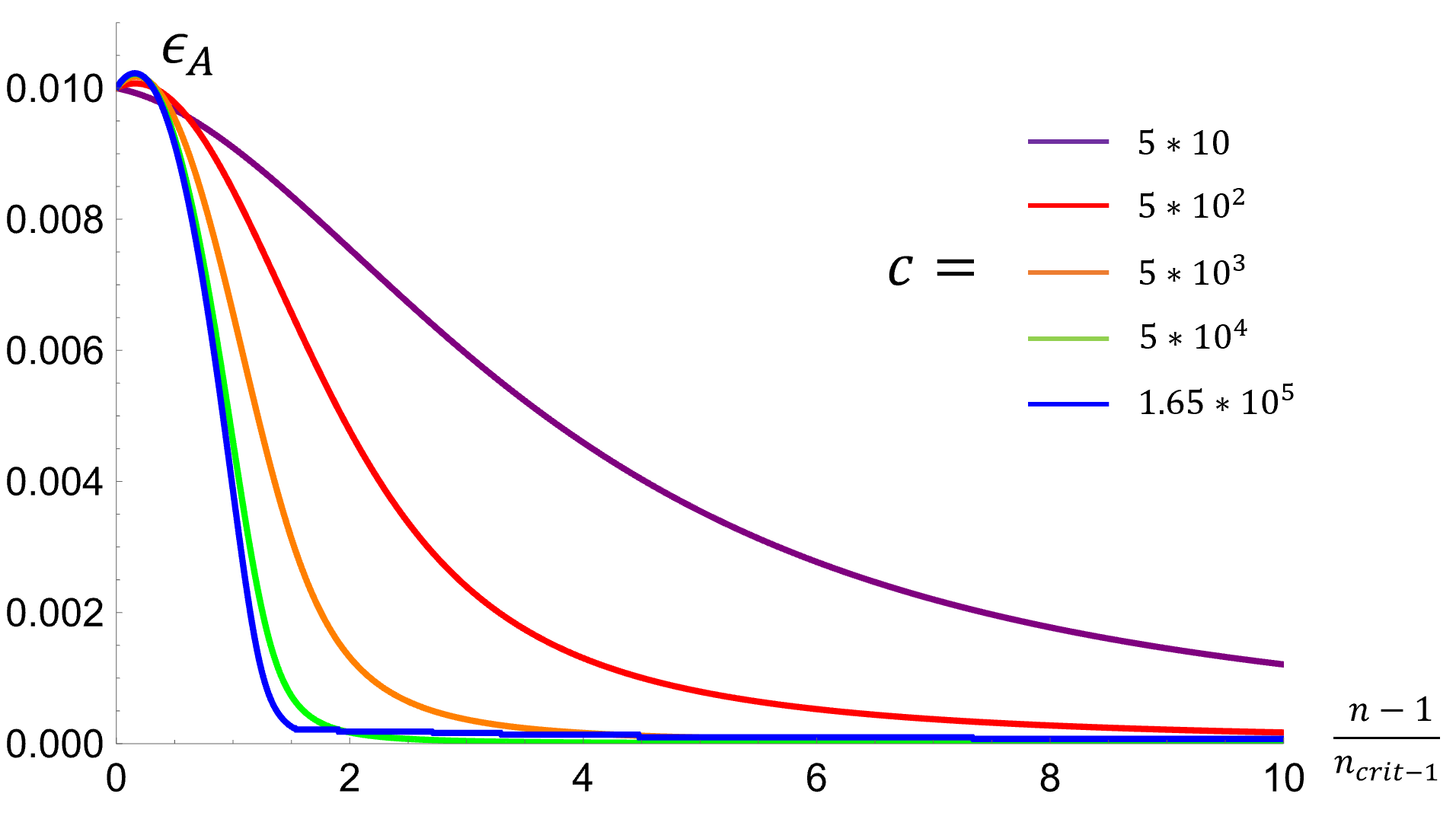}
    \includegraphics[width=0.48\linewidth]{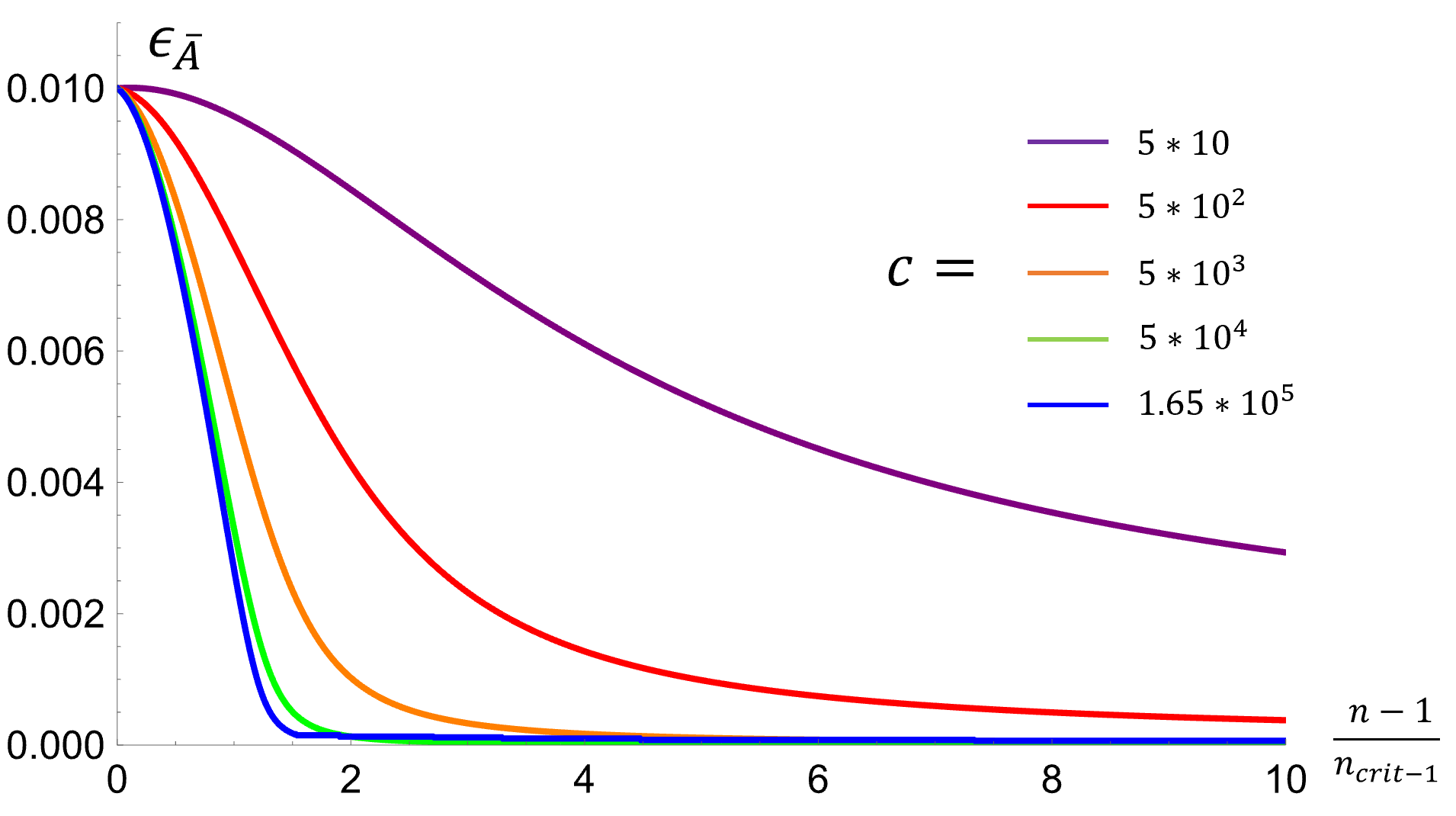}
    \includegraphics[width=0.48\linewidth]{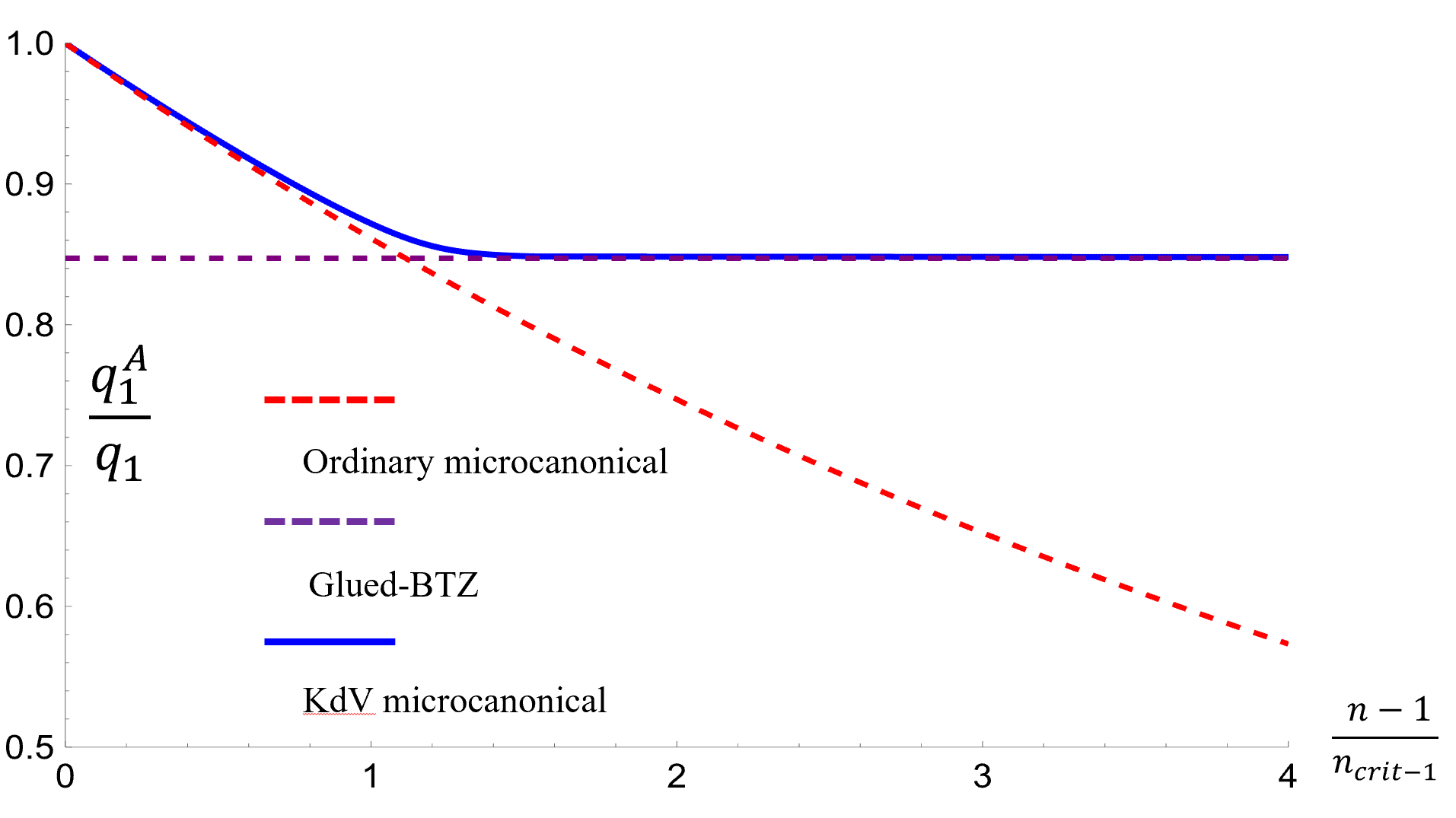}
    \includegraphics[width=0.48\linewidth]{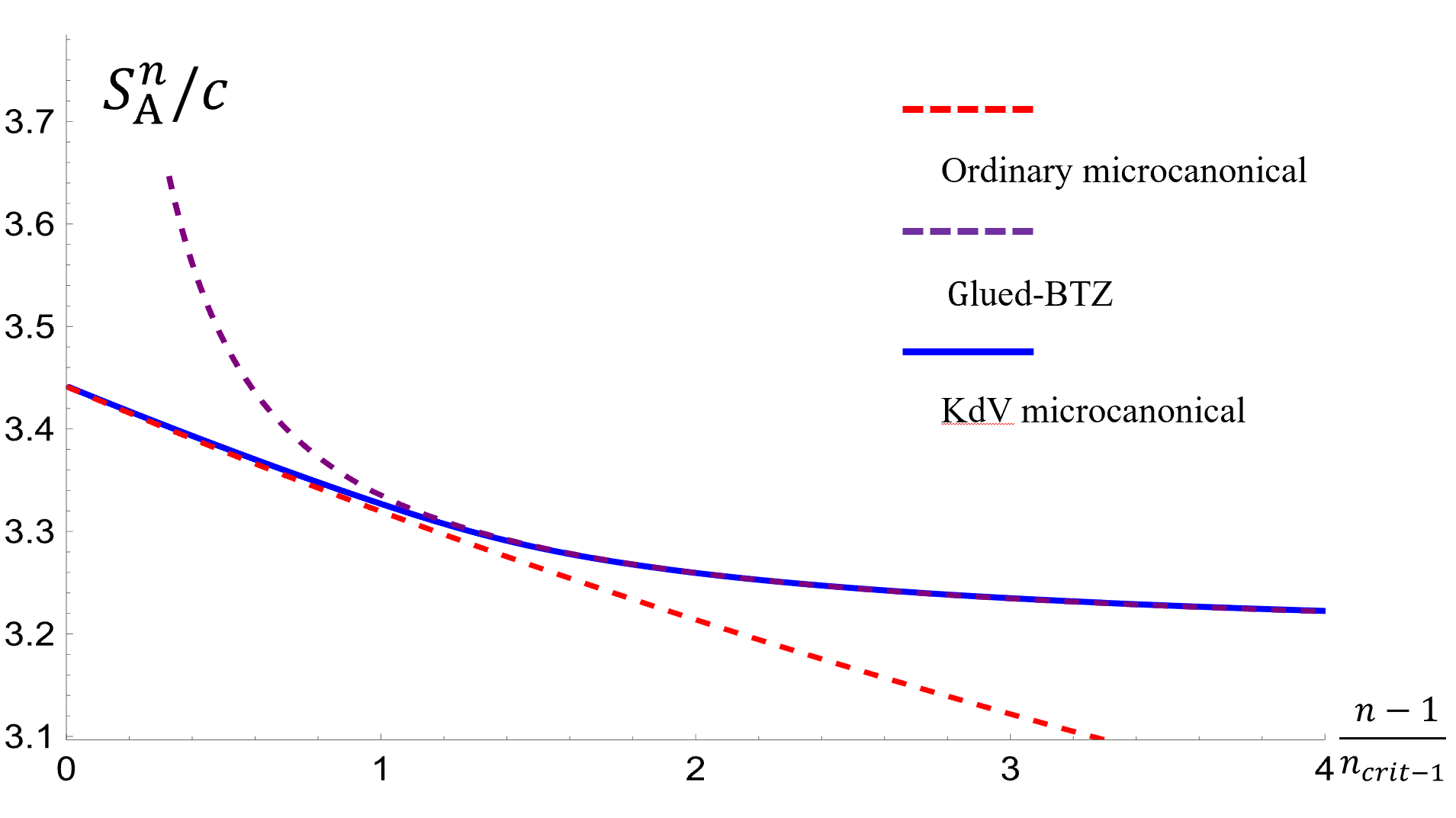}
    \caption{Top: index dependence of $\epsilon=q_3/q_1^2-1$ for the subsystem $A$ ($\epsilon^A$, left pandel) and its complement $\bar A$ ($\epsilon^{\bar A}$, right panel)  as a function of index $n$, for $c=\lbrace 50,5\cdot 10^2,5\cdot 10^3,5\cdot 10^4,1.65\cdot 10^5 \rbrace$. The crossover around  $n = n_{crit}$ becomes sharper in the holographic limit $c\to \infty$. Bottom: 
        the crossover behavior of the subsystem charge density $q^A_1/q_1$ (left) and subsystem Renyi entropy $S^n_A/c$ (right) as functions of index $n$, both shown in solid blue. The result is for parameters 
        $c=1.65\cdot 10^5, x=0.3$ and the KdV microcanonical ensemble with $ q_1 = 20 c, \epsilon = 10^{-2}$.
        For comparison we plot  the same quantities when the global state is the microcanonical ensemble with the same $q_1$ (dashed red), and for the holographic glued-BTZ configuration  \eqref{gbtz} (dashed purple).}
    \label{fig:transitions}
\end{figure}

In the holographic limit  $c\to \infty$ it can  be shown that $\kappa_0 \sim 1/c \to 0$. Therefore in the holographic limit one finds 
\be
\lim_{n\to \infty}\epsilon_A, \epsilon_{\bar{A}} \to 0, 
\ee
even when $\epsilon$ is finite.
In this limit the subsystem charge densities approach  $n$-independent values
\be
\label{gbtz}
q^A_1 = q_1 \left(1-\sqrt{\frac{(1-x)\epsilon}{x}}\right),\quad q^{\bar{A}}_1 = q_1 \left(1+\sqrt{\frac{x\epsilon}{1-x}}\right).
\ee
This corresponds to glued-BTZ solution, as was mentioned  in the main text (also see the companion paper \cite{KdVETHholography}). According to the discussion there, due to the property that at leading $1/c$ order $s(q_1,q_3)$ depends only on $q_1$, we expect to see a sharp transition between the behavior as in the case of the microcanonical ensemble (\ref{micro}) and that one associated with the holographic glued-BTZ solution (\ref{eq:Renyi_BTZ}). As a function of Renyi index $n$ the transition occurs at 
\be 
n_{crit} = \sqrt{q^{\bar{A}}_1/q^A_1},
\ee
with $q_1^A,q_1^{\bar A}$ given by \eqref{gbtz}.
To illustrate that, we show numerical results for $\epsilon^{A,\bar A}$ as a function of $n$ for different values of $c$, in Fig.~\ref{fig:transitions}, and observe that as $c$ increases the crossover region shrinks. We also plot numerical values of the subsystem Renyi entropy and ratio $q_1^A/q_1$ as functions of $n$ for numerically large $c=1.65\cdot 10^5$, in which case the transition region becomes very small. 
All numerical results are obtained with the approximate expression for the KdV microcanonical density of states \eqref{eq:KdV_entropy}.

\bibliographystyle{JHEP} 
\bibliography{ref}

\end{document}